\documentclass[10pt,onecolumn,english,prd,superscriptaddress,nofootinbib,preprintnumbers,showpacs,floatfix]{revtex4-2}
\usepackage{graphicx,color}
\usepackage{amsmath}
\usepackage{amssymb}
\usepackage{hyperref}
\usepackage[utf8]{inputenc}
\usepackage[english]{babel}
\usepackage{epsfig}
\usepackage{subfigure}
\usepackage{wasysym}
\usepackage{color,xcolor}
\usepackage{amsmath}
\usepackage{bm}
\usepackage{epsfig}
\usepackage{amsfonts}
\usepackage{dcolumn}
\usepackage{float}
\usepackage{makecell} 
\usepackage{comment}
\usepackage{cases}

\hypersetup{colorlinks=true, linkcolor=blue, citecolor=green}
\usepackage{dcolumn}
\usepackage{bm}
\usepackage{ifpdf}
\usepackage{hyperref}
\usepackage{float}
\usepackage{bm}
\usepackage{xcolor,color,graphicx,graphics}
\usepackage[OT1]{fontenc}
\usepackage{latexsym,amssymb,amsmath,amsfonts}
\usepackage{makeidx}
\usepackage{epsfig}
\usepackage{epstopdf}
\usepackage{mathrsfs}
\hypersetup{colorlinks=true, linkcolor=blue, citecolor=green}
\usepackage{enumerate}
\usepackage{xcolor}
\usepackage{multirow}
\usepackage{orcidlink}
\usepackage{placeins}

\usepackage{tikz}

\begin{document}

\title{Particle dynamics and quasi-periodic oscillations of a Reissner--Nordstr\"om-like black hole in Kalb--Ramond gravity under an external magnetic test field}

\author{Faizuddin Ahmed\orcidlink{0000-0003-2196-9622}}
\email{faizuddinahmed15@gmail.com}
\affiliation{Department of Physics, The Assam Royal Global University, Guwahati, 781035, Assam, India}

\author{Ahmad Al-Badawi\orcidlink{0000-0002-3127-3453}}
\email{ahmadbadawi@ahu.edu.jo}
\affiliation{Department of Physics, Al-Hussein Bin Talal University 71111, Ma'an, Jordan}

\author{Sardor~Murodov\orcidlink{0000-0003-2360-4475}}
\email{s.murodov@newuu.uz}
\affiliation{New Uzbekistan University, Movarounnahr Street 1, Tashkent 100000, Uzbekistan}
\affiliation{Institute of Fundamental and Applied Research, National Research University TIIAME, Kori Niyoziy 39, Tashkent 100000, Uzbekistan}

\author{Bekzod~Rahmatov\orcidlink{0009-0001-0394-650X}}
\email{rahmatovbekzod@samdu.uz}
\affiliation{University of Tashkent for Applied Sciences, Str. Gavhar 1, Tashkent 100149, Uzbekistan}
\affiliation{Tashkent State Technical University, Tashkent 100095, Uzbekistan}

\author{Javlon~Rayimbaev\orcidlink{0000-0001-9293-1838}}
\email{javlon@astrin.uz}
\affiliation{Institute of Theoretical Physics, National University of Uzbekistan, Tashkent 100174, Uzbekistan}

\date{\today}

\begin{abstract}
We investigate the dynamics of charged test particles and quasi-periodic oscillations around a Reissner--Nordstr\"om-like black hole in Kalb--Ramond (KR) gravity in the presence of an external magnetic test field. The KR background introduces a Lorentz-violating parameter $\ell$, which modifies the spacetime geometry, horizon structure, circular orbits, and characteristic frequencies of particle motion. In contrast to the standard Wald-type prescription, the magnetic-field configuration is constructed from the source-free Maxwell equation on the charged KR background, allowing the magnetic profile to be consistently adapted to the modified geometry. We derive the equations of motion, the effective potential, the conditions for circular orbits, and the orbital and radial epicyclic frequencies of charged particles. The results show that the black-hole charge $Q/M$, the KR parameter $\ell$, the specific particle charge $\epsilon$, and the magnetic coupling $\beta=bM$ jointly affect the innermost stable circular orbit (ISCO) and the quasi-periodic oscillation (QPO) frequencies. We then apply the obtained frequencies to the relativistic precession model, where the upper QPO frequency is identified with the orbital frequency and the lower one with the periastron-precession frequency. Using the observed twin-peak QPO data of GRO J1655--40, XTE J1550--564, and M82 X-1, we perform a Markov chain Monte Carlo analysis to constrain the model parameters. The obtained posterior constraints indicate that the charged KR black-hole model with an external magnetic field can consistently reproduce the observed QPO pairs within the adopted parameter ranges. These findings suggest that QPO observations may serve as a useful phenomenological tool for probing Lorentz-violating black-hole geometries and electromagnetic effects in strong-gravity environments.
\end{abstract}


\maketitle

\def\HMS{{\scriptscriptstyle{\rm HMS}}}

\section{Introduction}\label{sec1}

Black holes provide one of the most important natural laboratories for testing gravity in the strong-field regime. In general relativity, their exterior geometry is described by highly constrained solutions, but quantum-gravity inspired corrections, additional fields, and realistic astrophysical environments may produce small but observable deviations from the standard picture. Among these possibilities, Lorentz symmetry violation has received considerable attention because Lorentz invariance is a basic ingredient of both general relativity and quantum field theory, while several high-energy frameworks suggest that it may be an effective low-energy symmetry rather than an exact fundamental one. Early motivations came from string theory, where spontaneous Lorentz-symmetry breaking can arise through nonzero vacuum expectation values of tensor fields \cite{KosteleckySamuel1989,Samuel1989}. Related ideas have also been discussed in quantum-gravity phenomenology, noncommutative field theory, effective field theory, and cosmological settings \cite{AmelinoCamelia1998,Mattingly2005,Bertolami2005,Kanno2008,Carroll2004}. A systematic framework for parameterizing such effects is provided by the Standard-Model Extension and its gravitational sector \cite{ColladayKostelecky1998,Kostelecky2004,Kellie2021,Araújo2026}, while experimental and observational studies continue to place increasingly tight bounds on possible Lorentz-violating signals \cite{Abe2015}.

In gravitational physics, Lorentz-violating fields can modify compact-object spacetimes, particle motion, horizon structure, and wave propagation. Bumblebee-type models, in which a vector field acquires a nonzero vacuum expectation value, provide a useful example of spontaneous Lorentz symmetry breaking in curved spacetime \cite{Bluhm2005,Capelo2015,Casana2018}. Recent studies of black holes in Lorentz-violating and Bumblebee-inspired gravity, including analyses of Lorentz-violating electromagnetic and gravitational effects, have shown that such backgrounds may affect geodesic structure, lensing, shadows, thermodynamics, and other observable properties \cite{CarlosMarcos2025,Lessa2025,ZhuXu2025,Lai2026,MalufNeves2011}. These developments show that preferred vector or tensor fields can lead to measurable deviations not only in weak-field experiments, but also in the near-horizon region of compact objects.

Within this broader class of theories, the KR field is especially relevant. Originally introduced in string theory, the KR field is a rank-two antisymmetric tensor field $B_{\mu\nu}$ with field strength $H_{\mu\nu\rho}$ \cite{KalbRamond1974}. In gravitational contexts, this field is naturally related to torsion, axion-like degrees of freedom, and Lorentz-violating backgrounds. Static and spherically symmetric black-hole solutions in the presence of a background KR field have been studied in detail, showing that the corresponding metric can differ from the Schwarzschild or Reissner--Nordstr\"om forms through a Lorentz-violating parameter $\ell$ \cite{Yang2023,Duan2024,Ednaldo2024,EdnaldoJosé2024,Ednaldo2025,Daniela2025,DanielaEdnaldo2025}. Further investigations have considered the geodesic structure, deflection angle, thermodynamic behavior, and particle dynamics of KR and KR-ModMax black holes \cite{ALBADAWI2025102076,Badawiii,badawiA1,BadawiA2,Ahmed2026}. These works indicate that the KR parameter is not merely a formal deformation of the metric; it can change the horizon structure, the effective gravitational potential, and the characteristic frequencies of motion.

Several recent applications have connected KR gravity with astrophysical observables. Black holes surrounded by perfect fluid dark matter in KR gravity have been investigated from thermodynamic, dynamical, lensing, and QPO perspectives \cite{Jumaniyozov2025KRPFDM,Rahmatov2025KRPFDMPlasmaLensing}. The radiative properties and QPOs of charged black holes in KR gravity have also been analyzed \cite{Jumaniyozov2025KRChargedQPO}. More generally, compact-object tests in modified gravity have been extended to STVG black holes, quantum-corrected geometries, ModMax electrodynamics, scalar-hair black holes, phantom black holes, Simpson--Visser black holes, and other nonstandard backgrounds \cite{Rahmatov2026KSShadows,Rahmatov2026STVGLensing,Saydullayev2025STVPFDM,Rahmatov2026QuantumPlasmaLensing,Meliyeva2025ModMaxLensing,Zulqarnain2026ScalarHairOrbit,Shermatov2025PhantomQPO,Khan2026SimpsonVisserQPO,Nishonov2025STVGChargedQPO,Rahmatov2025ABGQPO,Murodov2026KSQPOThinDisk}. Related studies of wormholes, brane-world black holes, scalarized solutions, nonmetric stellar configurations, quark stars, and other compact objects further demonstrate the wide range of strong-field signatures that may arise beyond the standard black-hole paradigm \cite{Guo2026DeformedAdSTSW,Ditta2026TeleparallelWormhole,Errehymy2026HDEWormhole,Guo2026AcousticWH,Lutfuoglu2026BraneWorldQNM,Lutfuoglu2026ScalarizedBH,Khan2026NonmetricStellar,Banerjee2025QuarkStarsRainbow,Banerjee2025QCDEOSRainbow,Rahaman2008,Cheng2023,Farukh2024,Ashraf2025,Nishonov2025,Ahal2026,Meng2026}. These results motivate further study of KR-type black holes using observables that originate from the innermost regions of accretion flows.

QPOs are among the most promising probes of strong gravity. They are observed as nearly periodic modulations in the X-ray flux of accreting compact objects and are generally associated with dynamical processes in the inner accretion disk \cite{Psaltis2008,Bambi2012,Falanga2015,Belloni2012}. In particular, high-frequency QPOs are of special interest because their frequencies are comparable to orbital and epicyclic frequencies near the ISCO. The relativistic precession model and resonance-type models provide widely used phenomenological links between observed QPO pairs and characteristic frequencies of test-particle motion \cite{StellaVietri1998,AbramowiczKluzniak2001,Stuchlik2008,Stuchlik2013,Ditta2024}. Observationally, the microquasars GRO J1655--40 and XTE J1550--564, together with the intermediate-mass black-hole candidate M82 X-1, provide important QPO data sets for testing compact-object models \cite{Remillard1999GRO,Remillard2002HFQPO,Pasham2014M82}. Recent QPO studies in different black-hole backgrounds confirm that orbital and epicyclic frequencies can be sensitive to charge, deformation parameters, dark matter distributions, scalar fields, and other corrections to the spacetime geometry \cite{Wang2022,Liu2023,Sanjar2023,Boshkayev2023,Jumaniyozov2024,Guo2025,Jumaniyozov2025,Hazarika2025,Hazarika2026,Dasgupta2025,Shermatov2026DustQPO,Donmez2026JPAccretionQPO}.

Another essential element of realistic black-hole environments is the presence of magnetic fields. The study of black holes immersed in magnetic fields began with the classical Wald solution, and magnetic fields are now understood to be central to accretion dynamics, jet production, and electromagnetic energy extraction \cite{Wald1974,Wald1984book,BlandfordZnajek1977}. In the framework of black-hole electrodynamics and magnetohydrodynamics, magnetic fields can alter charged-particle trajectories, shift the ISCO, change epicyclic frequencies, and affect observable signatures of accretion flows \cite{AlievOzdemir2002,FrolovShoom2010,Zahrani2013,Kolos2015,Stuchlik2016,Tursunov2016,Gallego2020,Baker2023,Cao2024}. Polarimetric observations by the Event Horizon Telescope provide direct evidence that magnetic fields are important in the near-horizon plasma environment of supermassive black holes \cite{EHT2021}. Recent investigations of magnetized particles and magnetic-field configurations around black holes in modified gravity and nonstandard backgrounds further emphasize the importance of electromagnetic effects in strong-gravity tests \cite{Shermatov2026BraneworldMagnetic,Rahmatov2026JNWCurrentLoop,Murodov2025BBMBMagnetizedQPO,ALBADAWI2026170475}.

In the present work, we investigate the dynamics of charged particles around a Reissner--Nordstr\"om-like black hole in KR gravity in the presence of an external magnetic test field. A key point of our approach is that the magnetic field is not imposed through the usual Wald-type ansatz. Instead, it is obtained from the source-free Maxwell equation on the charged KR background. This is necessary because the KR parameter changes the asymptotic normalization of the spacetime, and therefore the magnetic profile must be adapted to the geometry. We derive the equations of motion, construct the effective potential, determine circular-orbit conditions, and analyze the ISCO behavior under the combined influence of the charge parameter $Q/M$, the Lorentz-violating parameter $\ell$, the specific particle charge $\epsilon$, and the magnetic coupling $\beta=bM$.

We then derive the orbital and radial epicyclic frequencies and apply them to the QPO problem within the relativistic precession model, where the upper frequency is identified with the orbital frequency and the lower frequency with the periastron-precession frequency. Finally, we perform a Markov chain Monte Carlo analysis using the \texttt{emcee} sampler \cite{ForemanMackey2013} and the observed QPO data of GRO J1655--40, XTE J1550--564, and M82 X-1. This statistical analysis allows us to constrain the model parameters and to examine whether the combined effects of the KR deformation and the external magnetic field can consistently reproduce the observed QPO pairs. In this way, the present study connects Lorentz-violating gravity, charged-particle dynamics, magnetic-field effects, and astrophysical QPO observations within a single phenomenological framework.

\section{Metric, electromagnetic potential and numerical magnetic-field solution}

We consider the static and spherically symmetric charged KR
black hole spacetime described by the line element
\begin{equation}
ds^{2}=-f(r)dt^{2}+f^{-1}(r)dr^{2}
+r^{2}\left(d\theta^{2}+\sin^{2}\theta\, d\phi^{2}\right),
\label{ds2}
\end{equation}
where the metric function is given by~\cite{Duan2024}
\begin{equation}
f(r)=\frac{1}{1-\ell}-\frac{2M}{r}
+\frac{Q^2}{(1-\ell)^2 r^2}.
\label{fr}
\end{equation}
Here $M$ is the mass parameter, $Q$ is the charge parameter of the charged KR
black hole, and $\ell$ is the Lorentz-violating parameter induced by the
KR field. In the limit $\ell\to0$, the metric function reduces to
the Reissner--Nordstr\"om form,
\begin{equation}
f(r)\Big|_{\ell=0}
=
1-\frac{2M}{r}+\frac{Q^2}{r^2}.
\end{equation}
For nonzero $\ell$, the spacetime is not asymptotically Minkowskian in the
standard coordinate normalization. Indeed,
\begin{equation}
f(r)\rightarrow \frac{1}{1-\ell},
\qquad
g_{rr}\rightarrow 1-\ell,
\qquad r\rightarrow\infty .
\label{asym_f}
\end{equation}
Thus the KR parameter modifies not only the asymptotic clock rate but also
the asymptotic radial normalization of the geometry. Throughout this work we
assume
\begin{equation}
\ell<1,
\label{ell_condition}
\end{equation}
so that the metric preserves the correct Lorentzian signature at large
distances.

The time coordinate normalized at spatial infinity is introduced as
\begin{equation}
t_{\rm phys}=\frac{t}{\sqrt{1-\ell}} .
\label{tphys}
\end{equation}
Therefore, the coordinate angular velocity
\begin{equation}
\Omega=\frac{d\phi}{dt}
\end{equation}
and the angular velocity measured with respect to the asymptotically
normalized time are related by
\begin{equation}
\Omega_{\rm phys}
=
\frac{d\phi}{dt_{\rm phys}}
=
\sqrt{1-\ell}\,\Omega .
\label{Omega_phys_def}
\end{equation}
Similarly, the conserved energy associated with the asymptotically normalized
time coordinate differs from the coordinate energy by the same normalization
factor,
\begin{equation}
E_{\rm phys}=\sqrt{1-\ell}\,E .
\label{Ephys_def}
\end{equation}
In the following derivations we use the coordinate-time conserved quantities
for the equations of motion, while observable angular frequencies are
converted to the asymptotically normalized form using Eq.~\eqref{Omega_phys_def}.

The horizon radii are obtained from $f(r)=0$ and are given by
\begin{equation}
r_{\pm}=(1-\ell)\left[
M\pm \sqrt{M^2-\frac{Q^2}{(1-\ell)^3}}
\right],
\label{horizons}
\end{equation}
where $r_{+}$ and $r_{-}$ denote the event and Cauchy horizons, respectively.
The existence of a black hole horizon requires
\begin{equation}
M^2\geq \frac{Q^2}{(1-\ell)^3},
\qquad\text{or equivalently}\qquad
|Q|\leq M(1-\ell)^{3/2}.
\label{horizon_condition}
\end{equation}
In the numerical analysis below we restrict attention to non-extremal black
holes,
\begin{equation}
|Q|<M(1-\ell)^{3/2},
\label{nonextremal_condition}
\end{equation}
for which $f'(r_+)\neq0$. The extremal case would require a separate
near-horizon expansion and is not considered in this work.

\begin{figure}[t]
\centering
\includegraphics[width=0.5\linewidth]{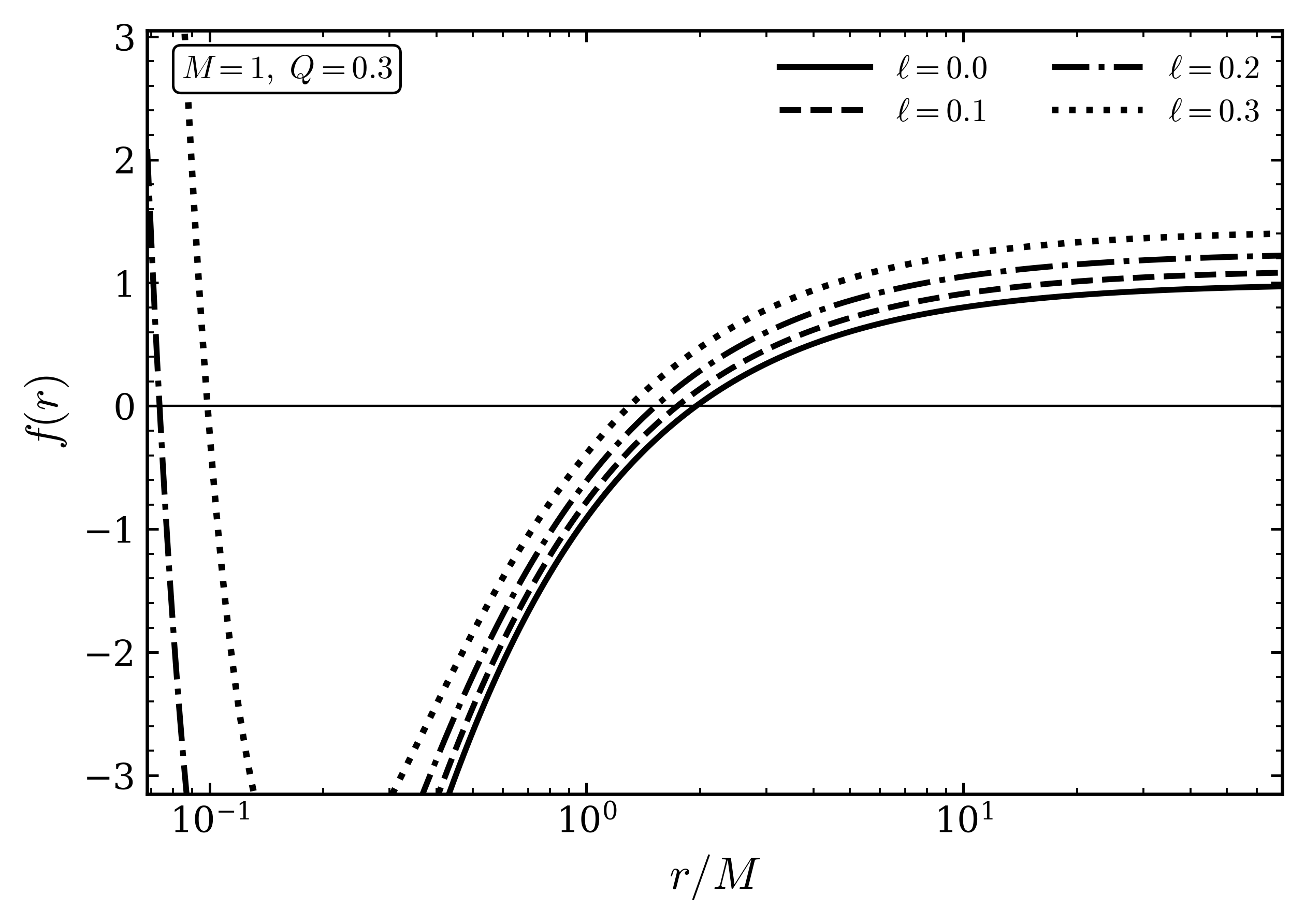}
\caption{Metric function $f(r)$ of the charged KR black hole for different
values of the Lorentz-violating parameter $\ell$. The parameters are fixed as
$M=1$ and $Q=0.3M$. The zero crossings determine the horizon positions, while
the large-radius value reflects the KR-modified asymptotic normalization.}
\label{fig:kr_metric_function}
\end{figure}

Figure~\ref{fig:kr_metric_function} illustrates how the KR parameter changes
the gravitational potential encoded in the metric function. The roots of
$f(r)$ determine the Cauchy and event horizons, while the large-radius value
$f(r)\to(1-\ell)^{-1}$ controls the asymptotic normalization of the time
coordinate. Thus, changing $\ell$ does not simply rescale the Reissner--
Nordstr\"om solution; it simultaneously modifies the near-horizon structure,
the effective strength of the charge term, and the normalization of the
exterior geometry. For fixed $M$ and $Q$, increasing $\ell$ raises the
asymptotic value of $f(r)$ and shifts the zero structure of the metric,
showing that the Lorentz-violating KR background changes both the local
gravitational redshift and the global radial scale of the spacetime.

\begin{table}[t]
\centering
\caption{Horizon structure and asymptotic magnetic exponent for the charged KR black hole. The calculations are performed for $M=1$ and $Q=0.3M$. The quantities $r_-$ and $r_+$ denote the Cauchy and event horizons, respectively, while $s_+$ determines the leading asymptotic behavior of the magnetic radial profile $\Psi_{\rm KR}(r)\sim r^{s_+}$.}
\label{tab:kr_horizon_structure}
\begin{tabular}{ccccc}
\hline
$\ell$ & $Q_{\max}/M$ & $r_-/M$ & $r_+/M$ & $s_+$ \\
\hline
0.0 & 1.000000 & 0.046061 & 1.953939 & 2.000000 \\
0.1 & 0.853815 & 0.057385 & 1.742615 & 1.931782 \\
0.2 & 0.715542 & 0.073708 & 1.526292 & 1.860147 \\
0.3 & 0.585662 & 0.098811 & 1.301189 & 1.784523 \\
\hline
\end{tabular}
\end{table}

The numerical values in Table~\ref{tab:kr_horizon_structure} provide a more
direct picture of this deformation. As $\ell$ grows from $0$ to $0.3$, the
event horizon moves inward, whereas the Cauchy horizon moves outward. The
separation between the two horizons therefore becomes smaller. At the same
time, the extremal charge allowed by the horizon condition,
$Q_{\max}=M(1-\ell)^{3/2}$, decreases. For a fixed charge $Q=0.3M$, the black
hole is consequently pushed closer to its extremal configuration as $\ell$
increases. This is physically important because the horizon structure controls
the redshift, the regularity condition for the magnetic-field equation, and
the innermost region accessible to stable particle motion. The decrease of
the exponent $s_+$ from its Reissner--Nordstr\"om value $s_+=2$ further shows
that the KR background weakens the standard asymptotic $r^2$ growth of a
uniform magnetic profile. Hence, the external magnetic test field must adapt
to the KR geometry rather than retain the usual Wald form.

\begin{figure}[t]
\centering
\includegraphics[width=0.5\linewidth]{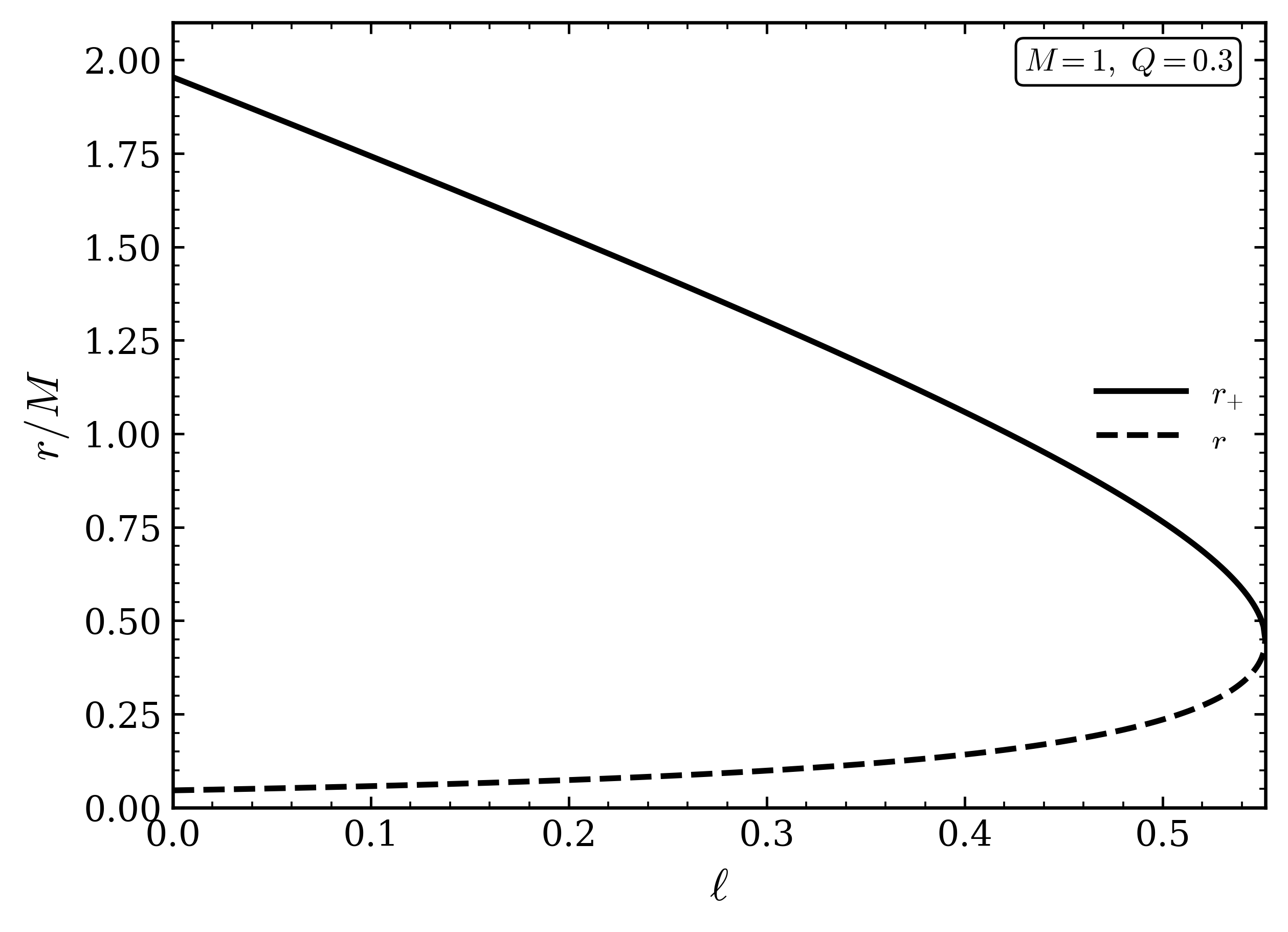}
\caption{Cauchy horizon $r_-$ and event horizon $r_+$ as functions of the KR
parameter $\ell$ for $M=1$ and $Q=0.3M$. Increasing $\ell$ decreases the
horizon separation and drives the fixed-charge configuration closer to the
extremal bound.}
\label{fig:kr_horizons}
\end{figure}

The electrostatic potential of the charged KR black hole is chosen as
\begin{equation}
A_t=-\Phi(r)=-\frac{Q}{(1-\ell)r}.
\label{At}
\end{equation}
With this convention, the radial electric-field component is
\begin{equation}
F_{rt}=\partial_r A_t
=
\frac{Q}{(1-\ell)r^2}.
\label{Frt}
\end{equation}

We now introduce an external magnetic test field aligned with the symmetry
axis of the black hole. The magnetic field is treated in the test-field
approximation: its energy density is assumed to be sufficiently small so that
the background KR geometry is not modified. Since the spacetime is static and
spherically symmetric, it admits the axial Killing vector
\begin{equation}
\xi^\mu_{(\phi)}
=
\left(\frac{\partial}{\partial \phi}\right)^\mu .
\end{equation}
For vacuum and asymptotically flat black hole backgrounds, a uniform magnetic
test field can often be introduced using the Wald prescription. However, in
the charged KR spacetime considered here, the simple choice
$A_\phi=(B/2)r^2\sin^2\theta$ is not, in general, an exact source-free
Maxwell solution. Moreover, for $\ell\neq0$ the asymptotic magnetic profile
does not have the standard $r^2$ behavior. Therefore, instead of imposing a
Wald-type radial profile, we determine the magnetic field directly from the
source-free Maxwell equation.

We take the electromagnetic four-potential in the form
\begin{equation}
A_\mu dx^\mu
=
A_t dt+A_\phi d\phi,
\qquad
A_\phi
=
\frac{B}{2}\Psi_{\rm KR}(r)\sin^2\theta ,
\label{Aphi_general}
\end{equation}
where $B$ is an overall magnetic-field amplitude and $\Psi_{\rm KR}(r)$ is an
unknown radial function. For $\ell\neq0$, $B$ should be understood as the
normalization amplitude of the KR-adapted magnetic test field, not as the
strength of an exactly uniform magnetic field at infinity. The nonzero
magnetic components of the electromagnetic tensor are
\begin{equation}
F_{r\phi}
=
\frac{B}{2}\Psi_{\rm KR}'(r)\sin^2\theta,
\qquad
F_{\theta\phi}
=
B\Psi_{\rm KR}(r)\sin\theta\cos\theta .
\label{magnetic_F}
\end{equation}
The source-free Maxwell equation
\begin{equation}
\nabla_\mu F^{\mu\nu}=0
\label{maxwell_eq}
\end{equation}
for $\nu=\phi$ gives the radial equation
\begin{equation}
\frac{d}{dr}
\left[
f(r)\frac{d\Psi_{\rm KR}(r)}{dr}
\right]
-
\frac{2\Psi_{\rm KR}(r)}{r^2}
=
0 .
\label{Psi_eq}
\end{equation}
Equivalently,
\begin{equation}
f(r)\Psi_{\rm KR}''(r)
+
f'(r)\Psi_{\rm KR}'(r)
-
\frac{2\Psi_{\rm KR}(r)}{r^2}
=
0 ,
\label{Psi_eq_expanded}
\end{equation}
where
\begin{equation}
f'(r)
=
\frac{2M}{r^2}
-
\frac{2Q^2}{(1-\ell)^2r^3}.
\label{fprime}
\end{equation}

The radial equation \eqref{Psi_eq} is homogeneous. Hence the overall
normalization of $\Psi_{\rm KR}(r)$ can be absorbed into the magnetic-field
amplitude $B$. We fix the physically relevant solution by imposing regularity
at the event horizon and by normalizing the growing asymptotic branch at large
radius.

For a non-extremal horizon, the regular near-horizon expansion is
\begin{equation}
\Psi_{\rm KR}(r)
=
\Psi_0+\Psi_1(r-r_+)
+\mathcal{O}\!\left[(r-r_+)^2\right],
\label{Psi_horizon_series}
\end{equation}
where substitution into Eq.~\eqref{Psi_eq} yields
\begin{equation}
\Psi_1
=
\frac{2\Psi_0}{r_+^2 f'(r_+)} .
\label{Psi1}
\end{equation}
Since Eq.~\eqref{Psi_eq} is linear, the initial value $\Psi_0$ may be chosen
arbitrarily. We first construct an unnormalized regular solution
$\widetilde{\Psi}(r)$ by setting
\begin{equation}
\widetilde{\Psi}(r_+)=1,
\qquad
\widetilde{\Psi}'(r_+)
=
\frac{2}{r_+^2 f'(r_+)} .
\label{Psi_initial}
\end{equation}
The integration is started slightly outside the event horizon,
\begin{equation}
r_{\rm in}=r_+ + \delta,
\qquad
\delta=10^{-6}M,
\label{rin_delta}
\end{equation}
using the expansion \eqref{Psi_horizon_series}. The ordinary differential
equation \eqref{Psi_eq_expanded} is then integrated outward with an adaptive
high-order Runge--Kutta method. In the numerical calculations all radii are
measured in units of $M$, and we use relative and absolute tolerances of
order $10^{-10}$. The outer integration radius is chosen as
\begin{equation}
R_\infty=10^4 M,
\label{Rinfty}
\end{equation}
and the stability of the extracted normalization is checked by increasing
$R_\infty$ and decreasing $\delta$.

At large distances, Eq.~\eqref{Psi_eq} reduces to
\begin{equation}
\frac{1}{1-\ell}\Psi_{\rm KR}''(r)
-
\frac{2\Psi_{\rm KR}(r)}{r^2}
\simeq 0 .
\end{equation}
Assuming a power-law behavior $\Psi_{\rm KR}\sim r^s$, one obtains
\begin{equation}
s(s-1)=2(1-\ell),
\end{equation}
and hence
\begin{equation}
s_{\pm}
=
\frac{1}{2}
\left[
1\pm\sqrt{9-8\ell}
\right].
\label{s_pm}
\end{equation}
The growing branch $s_+$ defines the KR-adapted asymptotic magnetic-field
profile. Since the numerical solution at a finite radius may contain both
asymptotic branches, we extract the leading coefficient by fitting
\begin{equation}
\widetilde{\Psi}(r)
=
\widetilde{C}_+ r^{s_+}
+
\widetilde{C}_- r^{s_-}
\label{asym_fit}
\end{equation}
over a large-radius interval, typically
$0.7R_\infty\leq r\leq R_\infty$. The normalized radial profile is then
defined by
\begin{equation}
\Psi_{\rm KR}(r)
=
\frac{M^{2-s_+}}{\widetilde{C}_+}
\widetilde{\Psi}(r).
\label{Psi_normalized}
\end{equation}
With this normalization,
\begin{equation}
\Psi_{\rm KR}(r)
\sim
M^{2-s_+}r^{s_+},
\qquad
r\rightarrow\infty .
\label{Psi_asym_normalized}
\end{equation}
Thus, for $\ell\neq0$, the magnetic field is a KR-adapted axial test field
rather than a strictly uniform magnetic field in the usual asymptotically flat
sense.

In the limit $\ell\to0$, one has $s_+\to2$, and the regular normalized
solution reduces to the Reissner--Nordstr\"om magnetic profile
\begin{equation}
\Psi_{\rm KR}(r)
\longrightarrow
r^2-Q^2 .
\label{RN_limit_Psi}
\end{equation}
For $Q=0$ this further reduces to the standard Wald-type Schwarzschild
profile $\Psi_{\rm KR}=r^2$. These limiting cases provide useful consistency
checks for the numerical construction.

The magnetic field measured by a local static orthonormal observer outside
the event horizon is obtained from the electromagnetic tensor. The nonzero
orthonormal components are
\begin{equation}
B^{\hat r}
=
\frac{B\Psi_{\rm KR}(r)}{r^2}\cos\theta,
\qquad
B^{\hat\theta}
=
-\frac{B\sqrt{f(r)}}{2r}
\Psi_{\rm KR}'(r)\sin\theta .
\label{B_tetrad}
\end{equation}
Here $B^{\hat r}$ describes the radial magnetic flux, while
$B^{\hat\theta}$ is the polar component measured in the local static frame.
The minus sign follows the standard orientation of an axial magnetic field,
$\mathbf{B}=B(\cos\theta\,\hat{r}-\sin\theta\,\hat{\theta})$, in the
asymptotically flat Schwarzschild limit. The local static frame is defined
only in the exterior region where $f(r)>0$.

\begin{figure}[t]
\centering
\includegraphics[width=0.5\linewidth]{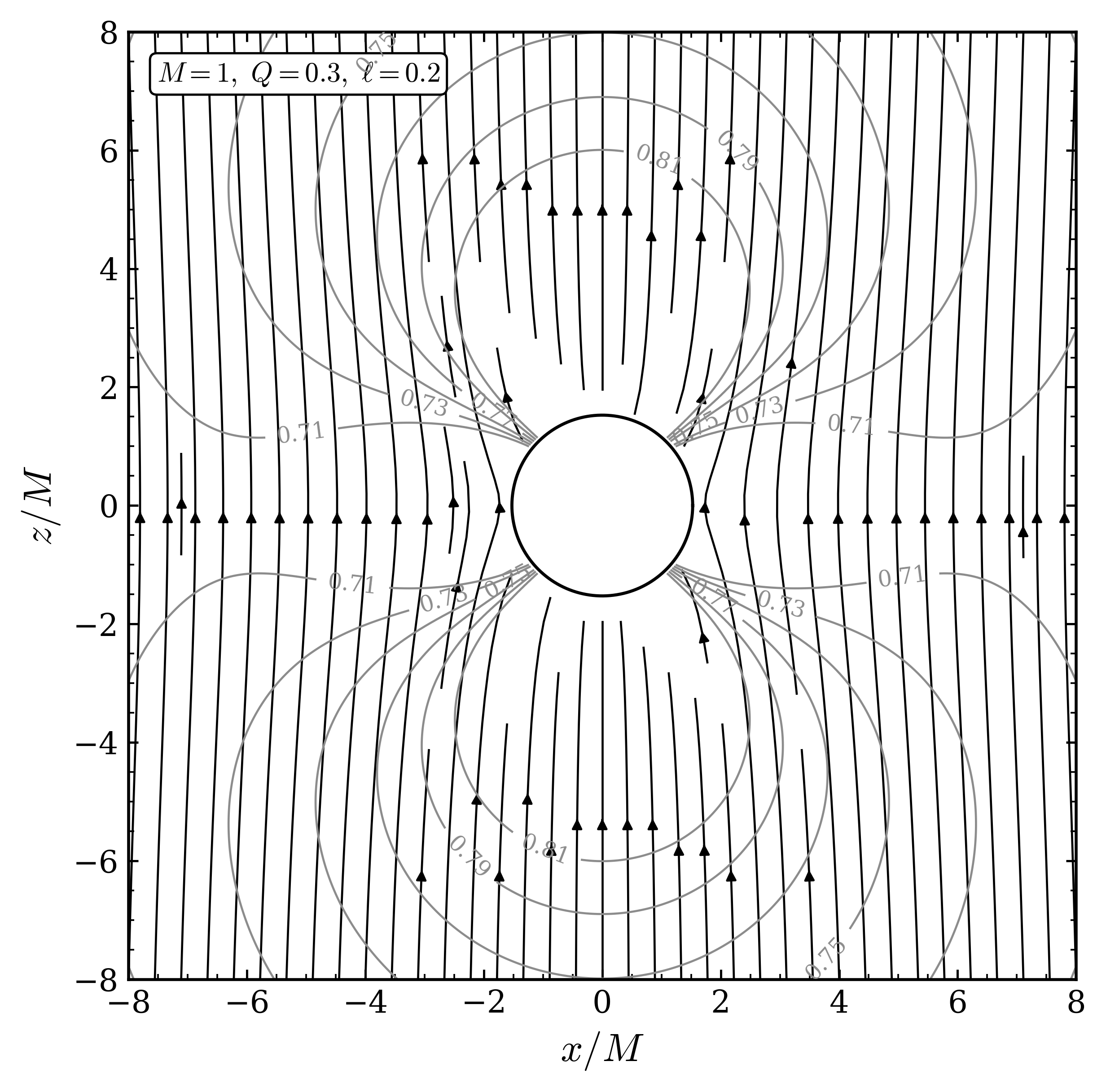}
\caption{Magnetic-field lines in the meridional plane for the charged KR black
hole. The parameters are $M=1$, $Q=0.3M$, and $\ell=0.2$. The circular black
curve represents the event horizon. The contours indicate representative
values of the local magnetic-field magnitude.}
\label{fig:kr_magnetic_field_lines}
\end{figure}

The corresponding magnetic-field configuration is shown in
Fig.~\ref{fig:kr_magnetic_field_lines}. The field lines are reconstructed from
the local orthonormal components in Eq.~\eqref{B_tetrad}. The radial component
is controlled by $\Psi_{\rm KR}(r)/r^2$, while the polar component is governed
by $\sqrt{f}\,\Psi'_{\rm KR}(r)/(2r)$. This separation is physically useful:
near the horizon the factor $\sqrt{f}$ suppresses the locally measured polar
component, whereas the radial flux is determined by the regular value of
$\Psi_{\rm KR}$. Far from the black hole, the field approaches the
KR-adapted power-law behavior fixed by $s_+$. Therefore, the bending and
redistribution of the field lines are not numerical artifacts; they reflect
how the charged KR geometry forces the source-free Maxwell solution to differ
from the standard uniform-field profile. The black circle in the figure marks
the event horizon, emphasizing that the plotted field is defined only in the
exterior static region.

Thus, in the present work, the magnetic field is not modeled by the simplified
Wald-type choice $\Psi_{\rm KR}=r^2$. Instead, the radial function
$\Psi_{\rm KR}(r)$ is obtained from the exact source-free Maxwell equation
\eqref{Psi_eq}, with regular boundary conditions at the event horizon and a
KR-adapted asymptotic normalization at large radius. This procedure provides
a self-consistent magnetic test-field configuration on the charged KR black
hole background.

\subsection{Lagrangian, equations of motion and conserved quantities}

The motion of a charged test particle of rest mass $m$ and charge $q$ is
governed by the Lagrangian
\begin{equation}
\mathscr{L}
=
\frac{1}{2}m\, g_{\mu\nu}\dot{x}^{\mu}\dot{x}^{\nu}
+qA_\mu\dot{x}^{\mu},
\label{lagrangian_general}
\end{equation}
where a dot denotes differentiation with respect to the proper time $\tau$.
The first term is the geodesic part, while the second describes the
electromagnetic interaction between the test particle and the background
electric and magnetic fields.

Substituting the metric and the electromagnetic potential, we obtain
\begin{equation}
\begin{split}
\mathscr{L}
=&\;
\frac{m}{2}\left[
-f(r)\dot{t}^{2}
+f^{-1}(r)\dot{r}^{2}
+r^{2}\dot{\theta}^{2}
+r^{2}\sin^{2}\theta\,\dot{\phi}^{2}
\right] \\
&+qA_t\dot{t}
+qA_\phi\dot{\phi}.
\end{split}
\label{lagrangian_explicit}
\end{equation}
The canonical momentum conjugate to $x^\mu$ is
\begin{equation}
p_\mu
=
\frac{\partial \mathscr{L}}{\partial \dot{x}^{\mu}}
=
mg_{\mu\nu}\dot{x}^{\nu}+qA_\mu.
\label{canonical_momentum}
\end{equation}
For a charged particle, the canonical momentum and the mechanical momentum
are therefore not identical.

Since the Lagrangian does not explicitly depend on $t$, the coordinate-time
energy is conserved:
\begin{equation}
E=-p_t=mf(r)\dot{t}-qA_t.
\label{energy_def}
\end{equation}
Introducing the specific coordinate energy and the specific charge-to-mass
ratio,
\begin{equation}
\mathcal{E}=\frac{E}{m},
\qquad
\varepsilon=\frac{q}{m},
\end{equation}
one obtains
\begin{equation}
\dot{t}
=
\frac{\mathcal{E}+\varepsilon A_t}{f(r)}.
\label{tdot}
\end{equation}
Outside the event horizon, $f(r)>0$, future-directed timelike motion requires
\begin{equation}
\mathcal{E}+\varepsilon A_t>0.
\label{future_condition}
\end{equation}
Since $A_t<0$ for $Q>0$ and $A_t>0$ for $Q<0$, the allowed motion depends on
both the sign of the black-hole charge and the sign of the test-particle
charge.

Since the Lagrangian does not explicitly depend on $\phi$, the axial angular
momentum is also conserved:
\begin{equation}
L=p_\phi
=
mr^2\sin^2\theta\,\dot{\phi}+qA_\phi .
\label{angular_momentum_def}
\end{equation}
Defining the specific angular momentum $\mathcal{L}=L/m$, we find
\begin{equation}
\dot{\phi}
=
\frac{\mathcal{L}-\varepsilon A_\phi}
{r^2\sin^2\theta}.
\label{phidot_general}
\end{equation}
The term $\varepsilon A_\phi$ shows that the magnetic field shifts the
mechanical angular velocity of the charged particle. Thus, even for fixed
canonical angular momentum $L$, the actual orbital angular velocity depends
on the magnetic interaction.

Because $A_\phi\propto\sin^2\theta$ is symmetric under the reflection
$\theta\to\pi-\theta$, the equatorial plane is invariant under the motion.
Thus the initial conditions
\begin{equation}
\theta=\frac{\pi}{2},
\qquad
\dot{\theta}=0
\end{equation}
ensure that the particle remains confined to the equatorial plane. We
therefore restrict the analysis to $\theta=\pi/2$.

On the equatorial plane,
\begin{equation}
A_\phi=\frac{B}{2}\Psi_{\rm KR}(r).
\label{Aphi_eq}
\end{equation}
It is convenient to introduce the magnetic coupling parameter
\begin{equation}
b=\frac{qB}{2m}=\frac{\varepsilon B}{2},
\label{b_def}
\end{equation}
which has the dimension of inverse length in geometrized units. For numerical
calculations we use the dimensionless magnetic parameter
\begin{equation}
\beta=bM=\frac{qBM}{2m}.
\label{beta_def}
\end{equation}
The parameter $b$ measures the strength and orientation of the magnetic
Lorentz interaction. Although $b$ and $\varepsilon Q$ both contain the same
charge-to-mass ratio $q/m$, they correspond to different physical effects:
$b$ controls the coupling to the external magnetic test field, whereas
$\varepsilon Q$ controls the electrostatic interaction with the charged KR
black hole. In parameter scans they may therefore be treated as independent
effective combinations when $B$ and $Q$ are varied independently.

Using Eq.~\eqref{Aphi_eq}, the azimuthal equation of motion becomes
\begin{equation}
\dot{\phi}
=
\frac{\mathcal{L}-b\Psi_{\rm KR}(r)}{r^2}.
\label{phidot_eq}
\end{equation}
Since $\Psi_{\rm KR}(r)$ is obtained numerically from the source-free Maxwell
equation, Eq.~\eqref{phidot_eq} is used directly in the particle-dynamics
calculations.

The coordinate angular velocity is
\begin{equation}
\Omega
=
\frac{d\phi}{dt}
=
\frac{
f(r)\left[\mathcal{L}-b\Psi_{\rm KR}(r)\right]
}{
r^2\left(\mathcal{E}+\varepsilon A_t\right)
}.
\label{Omega_coord}
\end{equation}
The angular velocity with respect to the asymptotically normalized time
$t_{\rm phys}$ is
\begin{equation}
\Omega_{\rm phys}
=
\sqrt{1-\ell}\,
\frac{
f(r)\left[\mathcal{L}-b\Psi_{\rm KR}(r)\right]
}{
r^2\left(\mathcal{E}+\varepsilon A_t\right)
}.
\label{Omega_phys_final}
\end{equation}

The normalization of the four-velocity,
\begin{equation}
g_{\mu\nu}\dot{x}^{\mu}\dot{x}^{\nu}=-1,
\label{normalization}
\end{equation}
gives, on the equatorial plane,
\begin{equation}
-f(r)\dot{t}^{2}
+f^{-1}(r)\dot{r}^{2}
+r^2\dot{\phi}^{2}
=-1.
\label{norm_equatorial}
\end{equation}
Substituting Eqs.~\eqref{tdot} and \eqref{phidot_eq} into
Eq.~\eqref{norm_equatorial} yields
\begin{equation}
\dot{r}^{2}
=
\left(\mathcal{E}+\varepsilon A_t\right)^2
-
f(r)\left[
1+
\frac{\left(\mathcal{L}-b\Psi_{\rm KR}(r)\right)^2}{r^2}
\right].
\label{radial_eq}
\end{equation}
We define the radial function
\begin{equation}
\mathcal{R}(r)
=
\left(\mathcal{E}+\varepsilon A_t\right)^2
-
f(r)\left[
1+
\frac{\left(\mathcal{L}-b\Psi_{\rm KR}(r)\right)^2}{r^2}
\right],
\label{R_def}
\end{equation}
so that
\begin{equation}
\dot{r}^{2}=\mathcal{R}(r).
\label{R_eq}
\end{equation}
The physically allowed region of motion is determined by
\begin{equation}
\mathcal{R}(r)\geq 0.
\label{allowed_region}
\end{equation}

The effective potential is obtained from the turning-point condition
$\dot{r}=0$. Solving Eq.~\eqref{radial_eq} for $\mathcal{E}$ gives two
branches,
\begin{equation}
V_{\rm eff}^{\pm}\!\left(r;\mathcal{L},b,\varepsilon,Q,\ell\right)
=
-\varepsilon A_t
\pm
\sqrt{
f(r)\left[
1+
\frac{\left(\mathcal{L}-b\Psi_{\rm KR}(r)\right)^2}{r^2}
\right]
}.
\label{Veff_pm}
\end{equation}
For future-directed motion outside the event horizon, the relevant branch is
\begin{equation}
V_{\rm eff}\!\left(r;\mathcal{L},b,\varepsilon,Q,\ell\right)
=
-\varepsilon A_t
+
\sqrt{
f(r)\left[
1+
\frac{\left(\mathcal{L}-b\Psi_{\rm KR}(r)\right)^2}{r^2}
\right]
}.
\label{Veff}
\end{equation}
Using Eq.~\eqref{At}, this becomes
\begin{equation}
V_{\rm eff}
=
\frac{\varepsilon Q}{(1-\ell)r}
+
\sqrt{
f(r)\left[
1+
\frac{\left(\mathcal{L}-b\Psi_{\rm KR}(r)\right)^2}{r^2}
\right]
}.
\label{Veff_final}
\end{equation}
The first term describes the electrostatic interaction between the particle
and the charged KR black hole. The square-root term encodes the gravitational
redshift, the rest-mass contribution, the centrifugal barrier, and the
magnetic correction to the mechanical angular momentum.

\begin{figure}[t]
\centering
\includegraphics[width=0.5\linewidth]{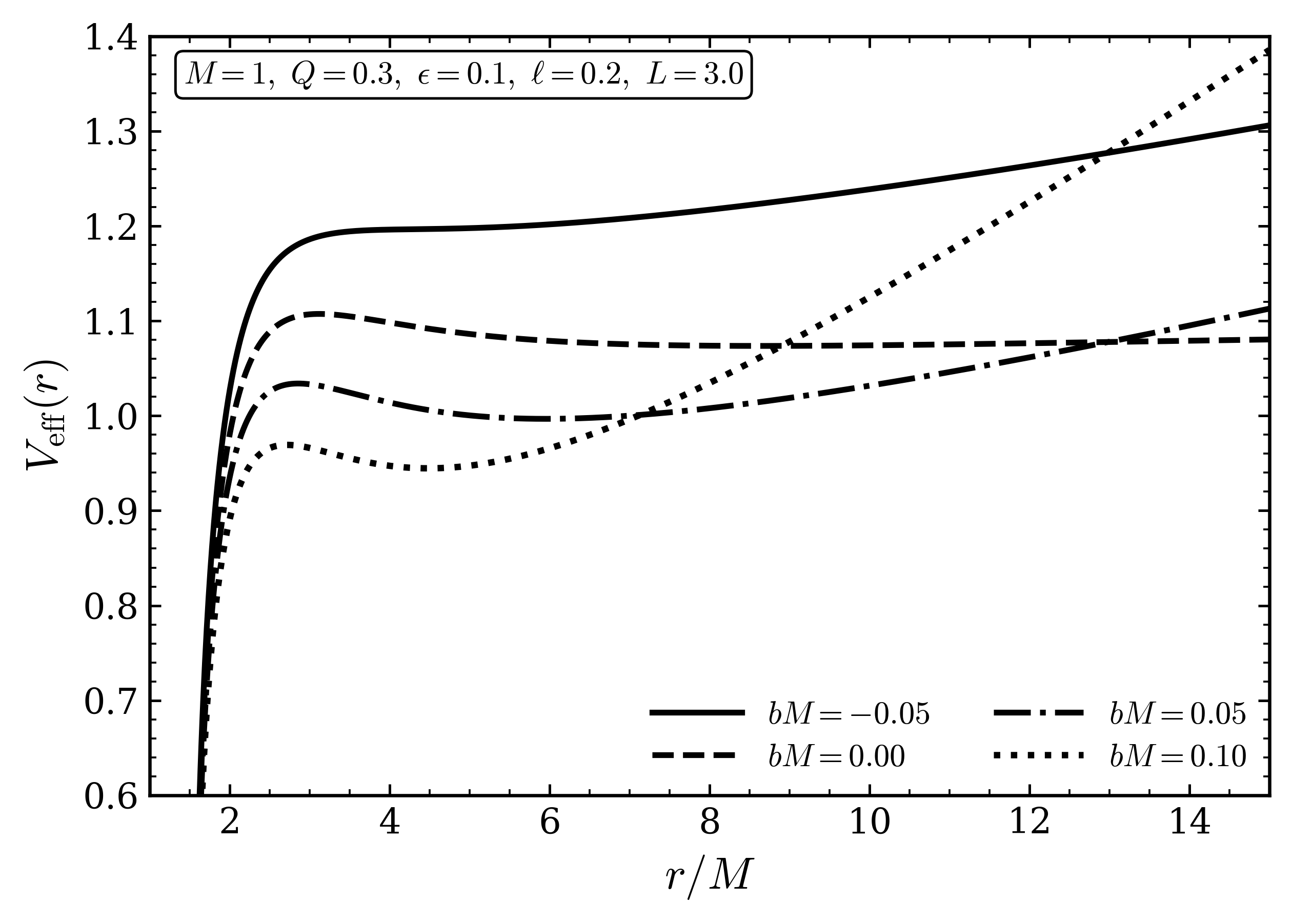}
\caption{Effective potential $V_{\rm eff}(r)$ for different magnetic coupling
parameters $bM$. The fixed parameters are $M=1$, $Q=0.3M$, $\varepsilon=0.1$,
$\ell=0.2$, and $\mathcal{L}=3M$. The variation with $bM$ comes from the
magnetic modification of the mechanical angular momentum
$\mathcal{L}-b\Psi_{\rm KR}$.}
\label{fig:kr_effective_potential}
\end{figure}

Figure~\ref{fig:kr_effective_potential} displays the effective potential for
different values of the magnetic coupling $bM$. The potential contains two
physically distinct contributions. The first term,
$\varepsilon Q/[(1-\ell)r]$, is the electrostatic interaction between the
charged particle and the charged KR black hole. For the parameters used in
the plot, $\varepsilon Q>0$, so this term is repulsive and is most important
close to the compact object. The second contribution is the square-root term,
which contains the gravitational redshift and the effective centrifugal
barrier through
$X(r)=\mathcal{L}-b\Psi_{\rm KR}(r)$. Since $\Psi_{\rm KR}(r)$ grows with
radius, even a small magnetic coupling can strongly modify the mechanical
angular momentum at large distances. Positive $bM$ reduces $X$ for fixed
canonical $\mathcal{L}$, while negative $bM$ increases it. The magnetic field
therefore changes the height and radial location of the potential barrier by
altering the balance between centrifugal support, electrostatic repulsion,
and gravitational attraction. The shift of the minimum of $V_{\rm eff}$
indicates that the preferred circular-orbit region is sensitive to the sign
and magnitude of the Lorentz coupling.

The turning points of the radial motion satisfy
\begin{equation}
\dot{r}=0,
\qquad
\mathcal{E}
=
V_{\rm eff}\!\left(r;\mathcal{L},b,\varepsilon,Q,\ell\right).
\label{turning_points}
\end{equation}
Circular orbits are determined by the simultaneous conditions
\begin{equation}
\mathcal{R}(r)=0,
\qquad
\frac{d\mathcal{R}(r)}{dr}=0,
\label{circular_conditions}
\end{equation}
where the derivative is taken with respect to $r$ at fixed conserved
quantities $\mathcal{E}$ and $\mathcal{L}$.

For compactness, define
\begin{equation}
X(r)=\mathcal{L}-b\Psi_{\rm KR}(r).
\label{X_def}
\end{equation}
Differentiating Eq.~\eqref{R_def} with respect to $r$ and setting the result
to zero gives
\begin{equation}
2\left(\mathcal{E}+\varepsilon A_t\right)\varepsilon A_t'
-
f'\left[
1+
\frac{X^2}{r^2}
\right]
+
\frac{2fbX\Psi_{\rm KR}'}{r^2}
+
\frac{2fX^2}{r^3}
=0.
\label{circular_explicit}
\end{equation}
The last two terms arise from differentiating $X^2/r^2$ at fixed
$\mathcal{L}$ and $b$:
\begin{equation}
\frac{d}{dr}\left(\frac{X^2}{r^2}\right)
=
-\frac{2bX\Psi_{\rm KR}'}{r^2}
-\frac{2X^2}{r^3}.
\end{equation}

For nonzero electric and magnetic couplings, the circular-orbit equations do
not generally admit a simple closed-form solution. In the numerical
implementation, for each chosen radius $r$ we substitute the physical
turning-point branch
\begin{equation}
\mathcal{E}
=
-\varepsilon A_t
+
\sqrt{
f(r)\left[
1+\frac{X^2}{r^2}
\right]
}
\label{Ecirc_general}
\end{equation}
into Eq.~\eqref{circular_explicit}. This gives a single nonlinear equation
for $\mathcal{L}(r)$:
\begin{equation}
2\varepsilon A_t'
\sqrt{
f(r)\left[
1+\frac{X^2}{r^2}
\right]
}
-
f'\left[
1+
\frac{X^2}{r^2}
\right]
+
\frac{2fbX\Psi_{\rm KR}'}{r^2}
+
\frac{2fX^2}{r^3}
=0.
\label{L_equation}
\end{equation}
After solving Eq.~\eqref{L_equation} numerically for $\mathcal{L}(r)$, the
corresponding specific energy is obtained from Eq.~\eqref{Ecirc_general}.
This procedure is equivalent to solving the coupled system
$\mathcal{R}=0$ and $\mathcal{R}'=0$ for
$\{\mathcal{E},\mathcal{L}\}$.

\begin{figure}[t]
\centering
\includegraphics[width=0.5\linewidth]{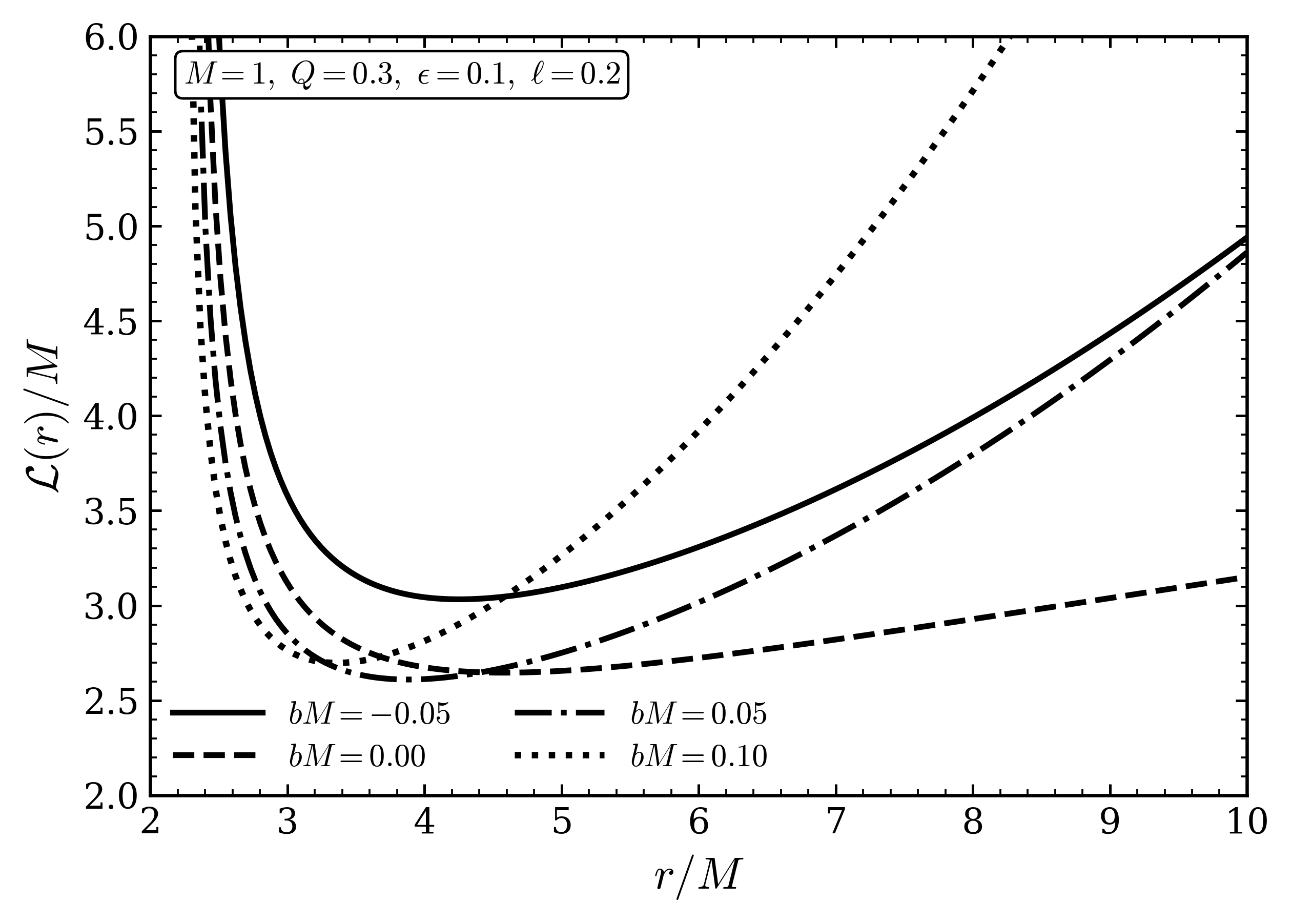}
\caption{Specific canonical angular momentum $\mathcal{L}(r)$ of
charged-particle circular orbits for different values of $bM$. The parameters
are $M=1$, $Q=0.3M$, $\varepsilon=0.1$, and $\ell=0.2$. The curve separation
reflects the magnetic vector-potential contribution to the canonical angular
momentum.}
\label{fig:kr_L_circular}
\end{figure}

\begin{figure}[t]
\centering
\includegraphics[width=0.5\linewidth]{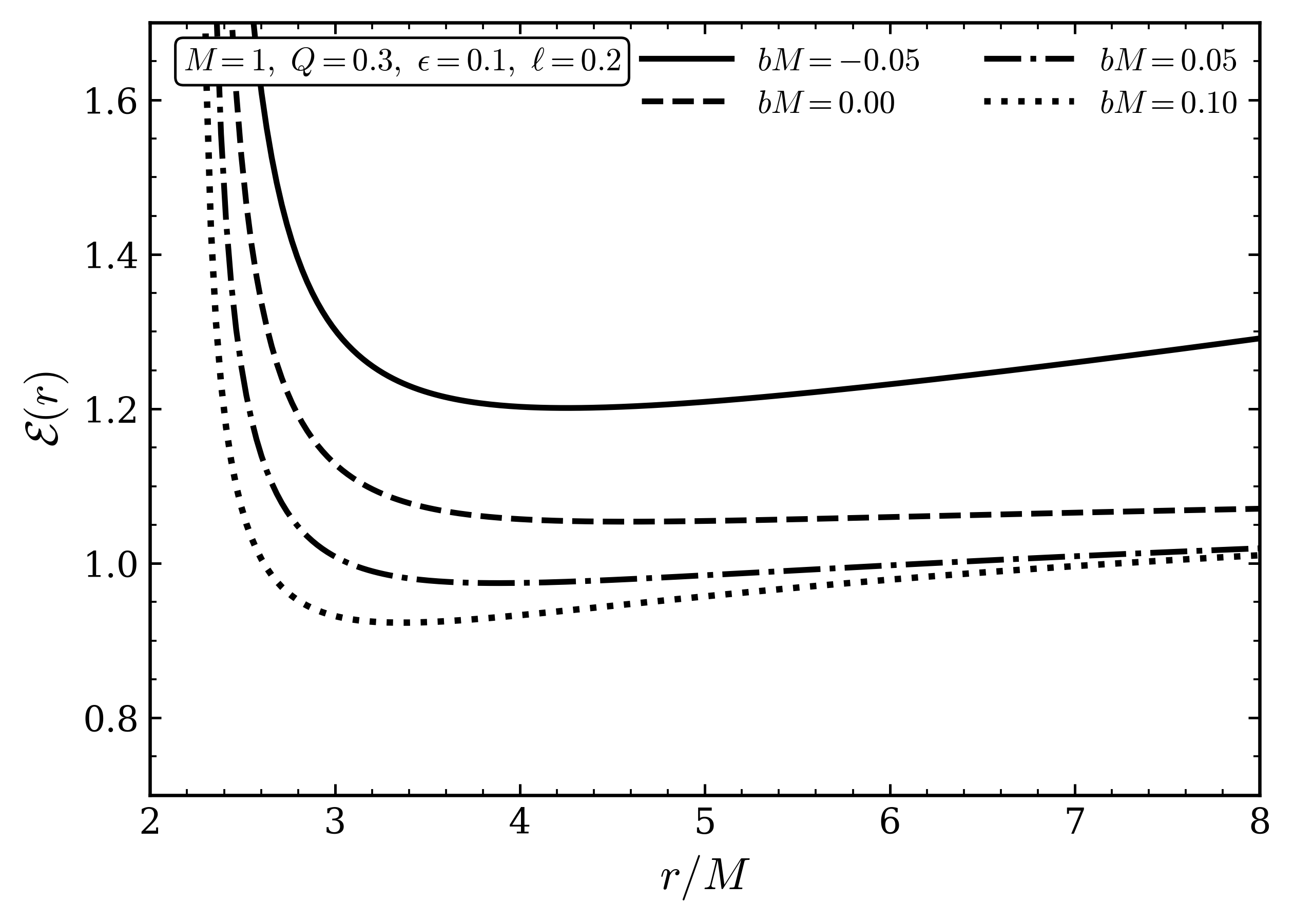}
\caption{Specific energy $\mathcal{E}(r)$ of charged-particle circular orbits
for different values of $bM$. The parameters are $M=1$, $Q=0.3M$,
$\varepsilon=0.1$, and $\ell=0.2$. The smoother response compared with
$\mathcal{L}(r)$ follows from the energy being controlled by the full
gravitational--electromagnetic balance.}
\label{fig:kr_E_circular}
\end{figure}

The circular-orbit values of the specific angular momentum and specific
energy are shown in Figs.~\ref{fig:kr_L_circular} and
\ref{fig:kr_E_circular}. These two quantities respond differently to the
magnetic field because the conserved angular momentum is canonical, whereas
the actual azimuthal motion is controlled by the mechanical combination
$X=\mathcal{L}-b\Psi_{\rm KR}$. For a fixed circular orbit, the radial force
balance first fixes the mechanical angular momentum needed to counteract
gravity and electrostatic repulsion. The canonical angular momentum must then
absorb the magnetic vector-potential contribution, so that
$\mathcal{L}=X+b\Psi_{\rm KR}$. This explains why the curves for
$\mathcal{L}(r)$ separate clearly as $bM$ changes. The separation becomes more
pronounced at larger radii because $\Psi_{\rm KR}(r)$ increases outward. By
contrast, the specific energy depends on the same magnetic correction only
through the square-root combination in Eq.~\eqref{Ecirc_general}; therefore
$\mathcal{E}(r)$ varies more smoothly. Physically, the magnetic field mainly
redistributes the angular-momentum budget of the orbit, while the energy is
controlled by the combined gravitational redshift, electrostatic interaction,
and kinetic support required for circular motion.

The stability of circular orbits can be analyzed either through the effective
potential or, equivalently, through the radial function. A stable circular
orbit corresponds to a local minimum of the effective potential,
\begin{equation}
\frac{d^2 V_{\rm eff}}{dr^2}>0.
\label{stability_V}
\end{equation}
The equivalent stability condition in terms of the radial function has the
opposite sign. Indeed,
\begin{equation}
\mathcal{R}(r)
=
\left(\mathcal{E}-V_{\rm eff}^{+}\right)
\left(\mathcal{E}-V_{\rm eff}^{-}\right),
\label{R_factorized}
\end{equation}
and for the physical branch $\mathcal{E}=V_{\rm eff}^{+}$ at a circular
orbit one obtains
\begin{equation}
\mathcal{R}''(r)
=
-
\left(\mathcal{E}-V_{\rm eff}^{-}\right)
\left(V_{\rm eff}^{+}\right)'',
\label{R_second_relation}
\end{equation}
where primes denote differentiation with respect to $r$ at fixed conserved
quantities. Since $\mathcal{E}-V_{\rm eff}^{-}>0$ outside the horizon, the
condition $V_{\rm eff}''>0$ is equivalent to
\begin{equation}
\frac{d^2\mathcal{R}}{dr^2}<0.
\label{stability_R}
\end{equation}
The marginally stable circular orbit is determined by
\begin{equation}
\frac{d^2 V_{\rm eff}}{dr^2}=0
\qquad\Longleftrightarrow\qquad
\frac{d^2\mathcal{R}}{dr^2}=0,
\label{marginal_stability}
\end{equation}
and corresponds to the ISCO in the appropriate
parameter range.

\begin{figure}[t]
\centering
\includegraphics[width=0.5\linewidth]{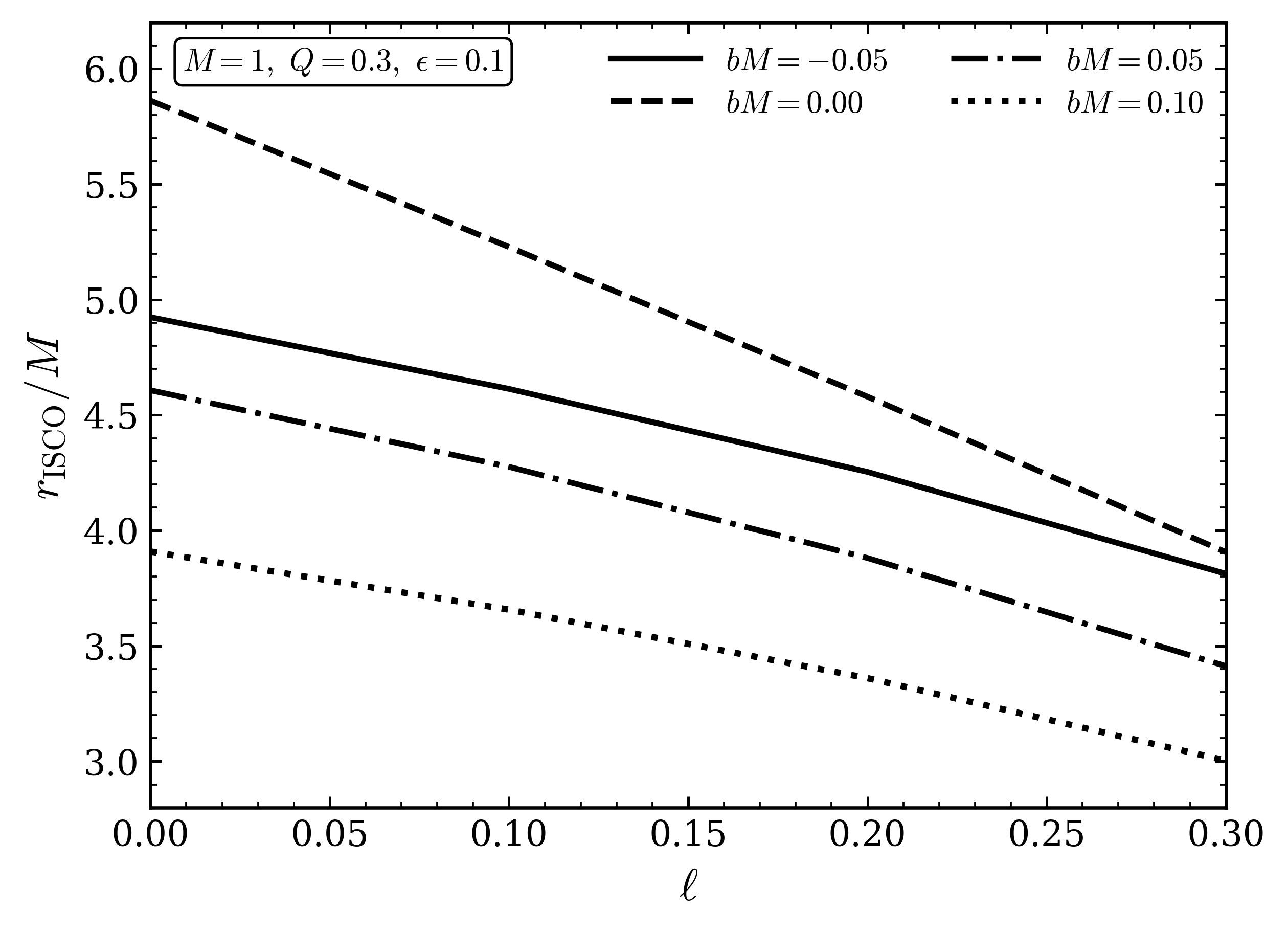}
\caption{ISCO radius as a function of the KR parameter $\ell$ for different
values of the magnetic coupling $bM$. The parameters are fixed as $M=1$,
$Q=0.3M$, and $\varepsilon=0.1$. The inward shift with increasing $\ell$
reflects the KR-induced change of the radial stability condition.}
\label{fig:kr_isco_ell}
\end{figure}

\begin{figure}[t]
\centering
\includegraphics[width=0.5\linewidth]{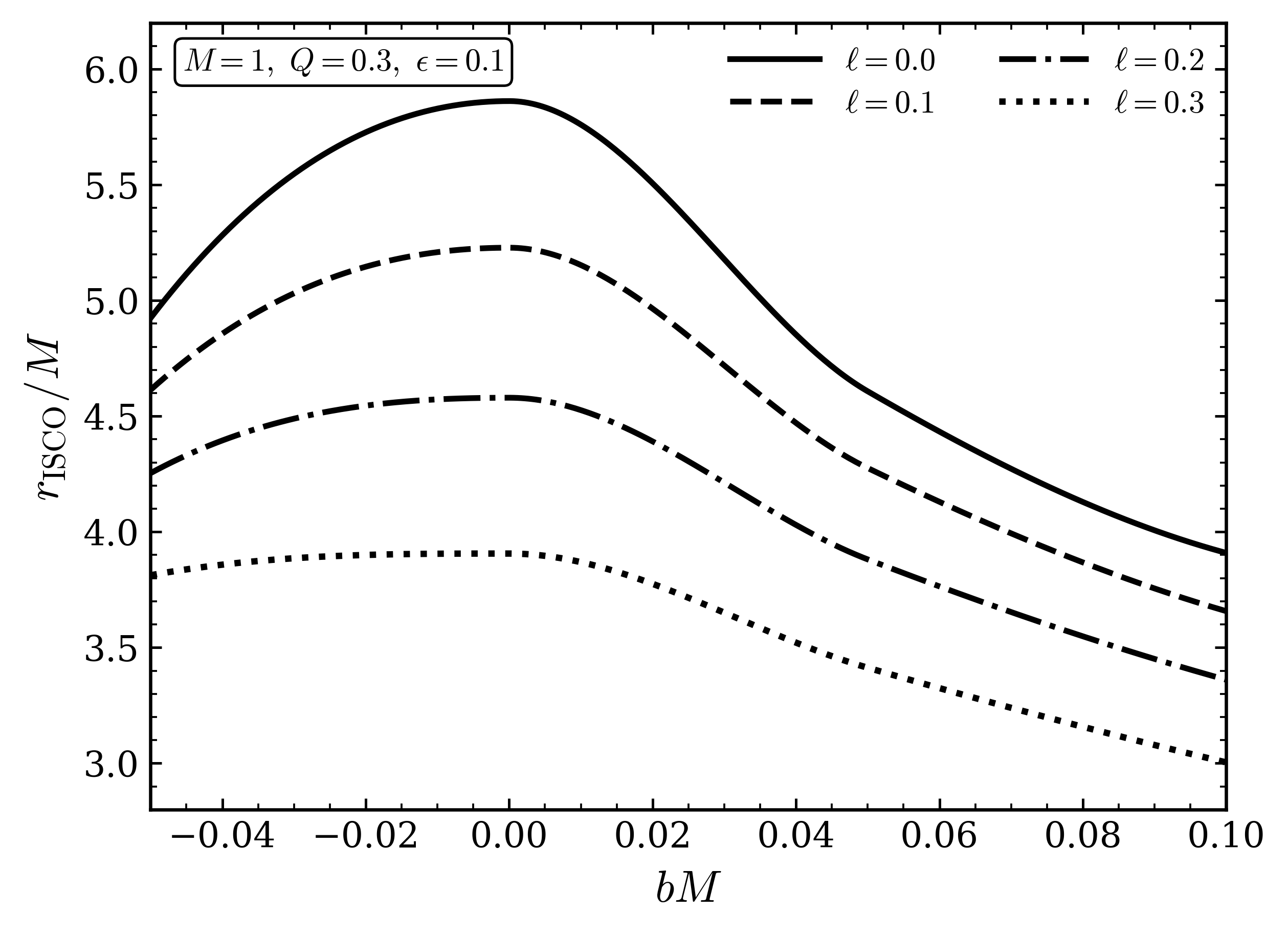}
\caption{ISCO radius as a function of the magnetic coupling parameter $bM$ for
different values of the KR parameter $\ell$. The magnetic interaction changes
the mechanical angular momentum and therefore shifts the marginally stable
orbit even though the magnetic field is treated as a test field.}
\label{fig:kr_isco_bM}
\end{figure}

The marginally stable orbit is obtained from Eq.~\eqref{marginal_stability}.
The dependence of $r_{\rm ISCO}$ on $\ell$ and $bM$ is summarized in
Figs.~\ref{fig:kr_isco_ell} and \ref{fig:kr_isco_bM}, with the corresponding
numerical values reported in Table~\ref{tab:kr_isco_quantities}. The ISCO is
a sensitive probe of the competition between gravity, electric repulsion,
centrifugal support, and the magnetic Lorentz interaction. Increasing $\ell$
changes the metric normalization and reduces the event-horizon radius for the
fixed charge used here. As a result, the stable-orbit region moves closer to
the black hole, and $r_{\rm ISCO}$ decreases in the parameter interval shown.
The magnetic coupling then adds a second control parameter. Because $b$ enters
through $X=\mathcal{L}-b\Psi_{\rm KR}$, it directly changes the mechanical
angular momentum that appears in the radial potential. In the present
parameter set, switching on the magnetic interaction shifts the ISCO inward
relative to the non-magnetized case, with the strongest inward shift occurring
for positive $bM$. This indicates that the Lorentz force effectively helps the
particle maintain stable circular motion at smaller radii by modifying the
centrifugal balance. Thus, the KR deformation fixes the background
gravitational scale, while the magnetic coupling controls how deeply charged
particles can remain on stable circular orbits within that background.

\begin{table}[t]
\centering
\caption{ISCO radius, specific energy, specific angular momentum, and physical angular velocity for selected values of the KR parameter and magnetic coupling. The parameters are fixed as $M=1$, $Q=0.3M$, and $\varepsilon=0.1$.}
\label{tab:kr_isco_quantities}
\begin{tabular}{cccccc}
\hline
$\ell$ & $bM$ & $r_{\rm ISCO}/M$ & $\mathcal{E}_{\rm ISCO}$ & $\mathcal{L}_{\rm ISCO}/M$ & $M\Omega_{\rm phys}$ \\
\hline
0.0 & -0.05 & 4.924024 & 1.187846 & 4.486104 & 0.118744 \\
0.0 & 0.00 & 5.861913 & 0.943281 & 3.370828 & 0.069162 \\
0.0 & 0.05 & 4.607413 & 0.843499 & 3.260279 & 0.070705 \\
0.0 & 0.10 & 3.908894 & 0.788564 & 3.408998 & 0.078291 \\
0.1 & -0.05 & 4.613557 & 1.184140 & 3.693319 & 0.120992 \\
0.1 & 0.00 & 5.228147 & 0.994118 & 3.011006 & 0.077514 \\
0.1 & 0.05 & 4.275870 & 0.904598 & 2.940426 & 0.078445 \\
0.1 & 0.10 & 3.657231 & 0.851334 & 3.055029 & 0.086191 \\
0.2 & -0.05 & 4.253802 & 1.201170 & 3.032794 & 0.124623 \\
0.2 & 0.00 & 4.579763 & 1.054007 & 2.646078 & 0.088475 \\
0.2 & 0.05 & 3.880964 & 0.974474 & 2.609944 & 0.089065 \\
0.2 & 0.10 & 3.360648 & 0.923260 & 2.697327 & 0.096709 \\
0.3 & -0.05 & 3.812564 & 1.239676 & 2.467302 & 0.131642 \\
0.3 & 0.00 & 3.906242 & 1.125832 & 2.272580 & 0.103792 \\
0.3 & 0.05 & 3.412712 & 1.055803 & 2.262862 & 0.104425 \\
0.3 & 0.10 & 3.003532 & 1.007004 & 2.328608 & 0.111728 \\
\hline
\end{tabular}
\end{table}

The numerical values emphasize the size of these effects. In the
non-magnetized case, the ISCO decreases from
$r_{\rm ISCO}/M\simeq 5.8619$ at $\ell=0$ to
$r_{\rm ISCO}/M\simeq 3.9062$ at $\ell=0.3$. This change is not only a
coordinate displacement; it reflects a genuine modification of the radial
stability condition caused by the KR deformation of $f(r)$. At fixed
$\ell=0.2$, the magnetic coupling further moves the ISCO from
$r_{\rm ISCO}/M\simeq 4.5798$ for $bM=0$ to
$r_{\rm ISCO}/M\simeq 3.3606$ for $bM=0.10$. The corresponding physical
angular velocity increases as the orbit moves inward, as expected for a
smaller stable circular radius. Therefore, even though the magnetic field is
treated as a test field and does not backreact on the metric, its Lorentz
force can significantly affect the particle-dynamical observables. This makes
$r_{\rm ISCO}$ and $M\Omega_{\rm phys}$ useful diagnostics for separating the
geometrical effect of the KR parameter from the electromagnetic effect of the
external magnetic field.

As a consistency check, in the neutral and non-magnetized limit
$\varepsilon=0$ and $b=0$, the equations reduce to the standard geodesic
relations
\begin{equation}
\dot{t}=\frac{\mathcal{E}}{f(r)},
\qquad
\dot{\phi}=\frac{\mathcal{L}}{r^2},
\end{equation}
and
\begin{equation}
\dot{r}^2
=
\mathcal{E}^2
-
f(r)\left(
1+\frac{\mathcal{L}^2}{r^2}
\right).
\label{neutral_radial}
\end{equation}
For neutral circular geodesics, one obtains
\begin{equation}
\mathcal{L}^2
=
\frac{r^3 f'}{2f-rf'},
\qquad
\mathcal{E}^2
=
\frac{2f^2}{2f-rf'}.
\label{EL_neutral}
\end{equation}
These expressions are recovered from the general charged-particle equations
when the electric and magnetic interactions are switched off.

Therefore, the KR parameter $\ell$, the black-hole charge parameter $Q$, the
particle's specific charge $\varepsilon$, and the magnetic coupling $b$
jointly modify the effective potential, the allowed region of motion, the
circular-orbit structure, and the stability properties of charged particles
around the charged KR black hole.

\section{Frequencies of quasi-periodic oscillations}
\label{sec4}

In this section, we derive the fundamental frequencies of charged-particle
motion around the charged KR black hole in the presence of the
source-free numerical magnetic field. These frequencies are then used for the
QPO analysis within the relativistic precession model.

We consider equatorial circular motion with four-velocity
\begin{equation}
    u^\mu=u^t(1,0,0,\Omega),
    \qquad
    \Omega=\frac{d\phi}{dt}.
    \label{u_circular_KR}
\end{equation}
The normalization condition $u^\mu u_\mu=-1$ gives
\begin{equation}
    u^t
    =
    \frac{1}{\sqrt{-g_{tt}-\Omega^2 g_{\phi\phi}}}
    =
    \frac{1}{\sqrt{f(r)-r^2\Omega^2}},
    \label{ut_KR}
\end{equation}
where the second equality is written on the equatorial plane
$\theta=\pi/2$. Thus, timelike circular motion requires
\begin{equation}
    f(r)-r^2\Omega^2>0 .
    \label{timelike_condition_KR}
\end{equation}

The orbital angular velocity is determined from the radial component of the
Lorentz-force equation,
\begin{equation}
    \frac{du^\alpha}{d\tau}
    +
    \Gamma^\alpha_{\mu\nu}u^\mu u^\nu
    =
    \varepsilon F^\alpha{}_\beta u^\beta.
    \label{lorentz_force_KR}
\end{equation}
For the charged KR spacetime with
\begin{equation}
    A_t=-\frac{Q}{(1-\ell)r},
    \qquad
    A_\phi=\frac{B}{2}\Psi_{\rm KR}(r)\sin^2\theta,
\end{equation}
one has, on the equatorial plane,
\begin{equation}
    F_{rt}=\frac{Q}{(1-\ell)r^2},
    \qquad
    F_{r\phi}=\frac{B}{2}\Psi_{\rm KR}'(r).
\end{equation}
With the magnetic coupling parameter $b$, the radial force-balance equation becomes
\begin{equation}
    -f'(r)+2r\Omega^2
    =
    -2\sqrt{f(r)-r^2\Omega^2}
    \left[
        \frac{\varepsilon Q}{(1-\ell)r^2}
        +
        b\Omega\Psi_{\rm KR}'(r)
    \right].
    \label{Omega_balance_KR}
\end{equation}
Therefore, the orbital frequency $\Omega_\phi$ is obtained as the physical
root of
\begin{equation}
    \mathcal{F}_\Omega(r,\Omega)
    \equiv
    -f'(r)+2r\Omega^2
    +
    2\sqrt{f(r)-r^2\Omega^2}
    \left[
        \frac{\varepsilon Q}{(1-\ell)r^2}
        +
        b\Omega\Psi_{\rm KR}'(r)
    \right]
    =0 .
    \label{Omega_root_KR}
\end{equation}
The physical branch is chosen so that it reduces to the neutral Keplerian
frequency when $\varepsilon=b=0$:
\begin{equation}
    \Omega_\phi^2
    \longrightarrow
    \Omega_0^2
    =
    \frac{f'(r)}{2r}
    =
    \frac{M}{r^3}
    -
    \frac{Q^2}{(1-\ell)^2r^4}.
    \label{Omega_neutral_KR}
\end{equation}
For nonzero electric and magnetic couplings, Eq.~\eqref{Omega_root_KR}
contains both the Coulomb contribution from the charged KR black hole and the
Larmor-type correction generated by the external magnetic field. In what
follows, the coordinate-time orbital frequency is denoted by
$\Omega_K\equiv\Omega_\phi$. The corresponding physically normalized
frequency is obtained later by the same asymptotic time normalization used in
Eq.~\eqref{Omega_phys_KR}. This convention keeps the notation consistent with
the previous subsection, where coordinate-time conserved quantities are used
in the equations of motion and physical frequencies are obtained only after
the rescaling $t_{\rm phys}=t/\sqrt{1-\ell}$.

\begin{figure}[t]
\centering
\includegraphics[width=0.5\linewidth]{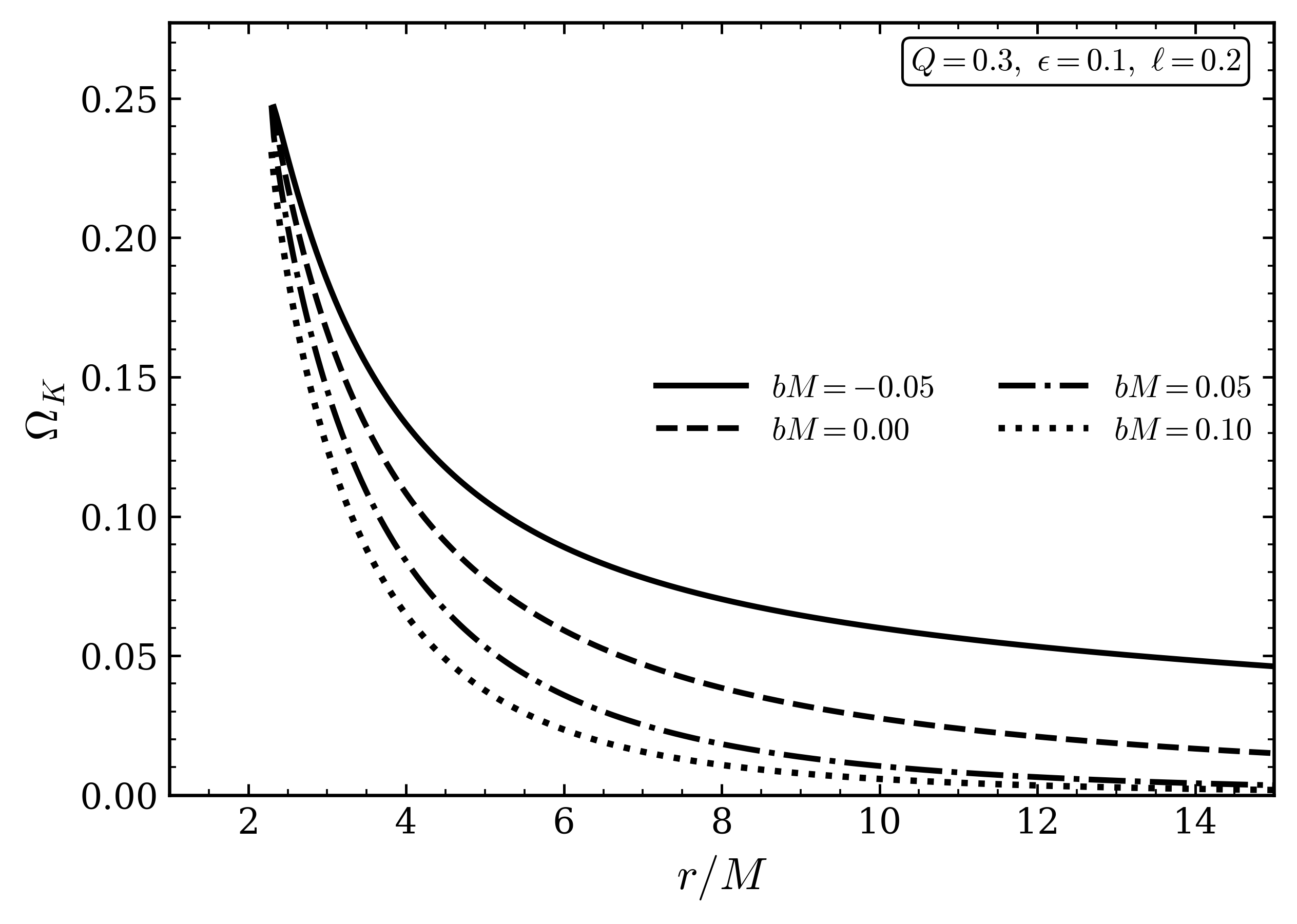}
\caption{Generalized Keplerian frequency $\Omega_K\equiv\Omega_\phi$ as a
function of radius for the charged KR black hole in the presence of
the source-free magnetic field. The parameters are fixed as
$M=1$, $Q=0.3M$, $\varepsilon=0.1$, and $\ell=0.2$, while several
values of the magnetic coupling $bM$ are shown.}
\label{fig:KR_OmegaK}
\end{figure}

Figure~\ref{fig:KR_OmegaK} shows the radial behavior of the generalized
Keplerian frequency. In the plot we use the compact label $\Omega_K$; the
physical frequency measured by an asymptotic observer is obtained through the
normalization $\Omega_K^{\rm phys}=\sqrt{1-\ell}\,\Omega_K$. The overall
decrease of $\Omega_K$ with increasing radius is a direct consequence of the
weakening gravitational attraction far from the compact object. Close to the
black hole, however, the orbital frequency is noticeably modified by the
combined action of the electric and magnetic interactions. The electric charge
of the spacetime enters through the Coulomb term
$\varepsilon Q/[(1-\ell)r^2]$, whereas the magnetic field changes the radial
force balance through the term $b\Omega_K\Psi'_{\rm KR}(r)$. Since
$\Psi_{\rm KR}(r)$ is the numerical solution of the source-free Maxwell
equation in the KR geometry, the magnetic correction is not an ad hoc shift
but a self-consistent response of the circular dynamics to the external field.
Thus the azimuthal period of a charged particle near the compact object is
determined not only by gravity, but also by the interplay between Coulomb and
Lorentz forces.

The specific energy and angular momentum of the circular orbit are then
written as
\begin{equation}
    \mathcal{E}
    =
    f(r)u^t-\varepsilon A_t,
    \qquad
    \mathcal{L}
    =
    r^2\Omega_\phi u^t+b\Psi_{\rm KR}(r).
    \label{EL_circular_frequency_KR}
\end{equation}

The radial and vertical epicyclic frequencies are obtained by perturbing the
motion around the circular orbit. To avoid confusion with the energy effective
potential $V_{\rm eff}$ introduced in the previous subsection, we denote the
auxiliary epicyclic potential by $\mathcal{U}(r,\theta)$:
\begin{equation}
    \mathcal{U}(r,\theta)
    =
    1
    +
    \frac{\left(\mathcal{E}+\varepsilon A_t\right)^2}{g_{tt}}
    +
    \frac{\left(\mathcal{L}-\varepsilon A_\phi\right)^2}{g_{\phi\phi}} .
    \label{U_epicyclic_KR}
\end{equation}
This quantity is not a second definition of $V_{\rm eff}$; it is only a
compact form of the four-velocity normalization used for deriving the
epicyclic frequencies. On the equatorial plane it is equivalent to the radial
function $\mathcal{R}(r)$ introduced earlier, up to the same circular-orbit
conditions. At a circular orbit,
\begin{equation}
    \mathcal{U}=0,
    \qquad
    \partial_r\mathcal{U}=0,
    \qquad
    \partial_\theta\mathcal{U}=0 .
\end{equation}

In Eq.~\eqref{U_epicyclic_KR}, the same specific quantities
$\mathcal{E}$, $\mathcal{L}$, $\varepsilon=q/m$, and
$b=\varepsilon B/2$ are used as in the previous subsection. In particular,
$\varepsilon A_\phi=b\Psi_{\rm KR}(r)\sin^2\theta$, and on the equatorial
plane this reduces to $\varepsilon A_\phi=b\Psi_{\rm KR}(r)$. Therefore the
mechanical angular-momentum combination is consistently
\[
    X(r)=\mathcal{L}-b\Psi_{\rm KR}(r),
\]
the same quantity that appears in the effective potential
$V_{\rm eff}$ and in the radial function $\mathcal{R}(r)$.
Small perturbations,
\begin{equation}
    r=r_0+\delta r,
    \qquad
    \theta=\frac{\pi}{2}+\delta\theta,
\end{equation}
satisfy
\begin{equation}
    \frac{d^2\delta r}{dt^2}+\Omega_r^2\delta r=0,
    \qquad
    \frac{d^2\delta\theta}{dt^2}+\Omega_\theta^2\delta\theta=0 .
\end{equation}
The epicyclic frequencies measured with respect to the coordinate time $t$
are
\begin{equation}
    \Omega_i^2
    =
    \frac{1}{2(u^t)^2g_{ii}}
    \frac{\partial^2\mathcal{U}}{\partial x_i^2}
    \bigg|_{r=r_0,\,\theta=\pi/2},
    \qquad
    x_i=(r,\theta).
    \label{epicyclic_general_KR}
\end{equation}

For the present KR geometry, the radial epicyclic frequency becomes
\begin{equation}
\begin{split}
    \Omega_r^2
    =
    f(r)
    \Bigg[
    &-
    \frac{f'(r)^2}{f(r)}
    +
    \frac{1}{2}f''(r)
    +
    3\Omega_\phi^2
    \\
    &-
    \frac{\varepsilon}{u^t}
    \left(
        A_t''(r)
        -
        2A_t'(r)\frac{f'(r)}{f(r)}
    \right)
    -
    \frac{\varepsilon^2 A_t'(r)^2}{(u^t)^2f(r)}
    \\
    &-
    \frac{b\Omega_\phi}{u^t}
    \left(
        \Psi_{\rm KR}''(r)
        -
        \frac{4\Psi_{\rm KR}'(r)}{r}
    \right)
    +
    \frac{b^2\Psi_{\rm KR}'(r)^2}{(u^t)^2r^2}
    \Bigg],
\end{split}
    \label{Omega_r_KR}
\end{equation}
where
\begin{equation}
    A_t'(r)=\frac{Q}{(1-\ell)r^2},
    \qquad
    A_t''(r)=-\frac{2Q}{(1-\ell)r^3}.
    \label{At_derivatives_KR}
\end{equation}
The magnetic terms contain $\Psi_{\rm KR}'(r)$ and
$\Psi_{\rm KR}''(r)$, which are evaluated from the numerical solution of
the source-free Maxwell equation.

\begin{figure}[t]
\centering
\includegraphics[width=0.5\linewidth]{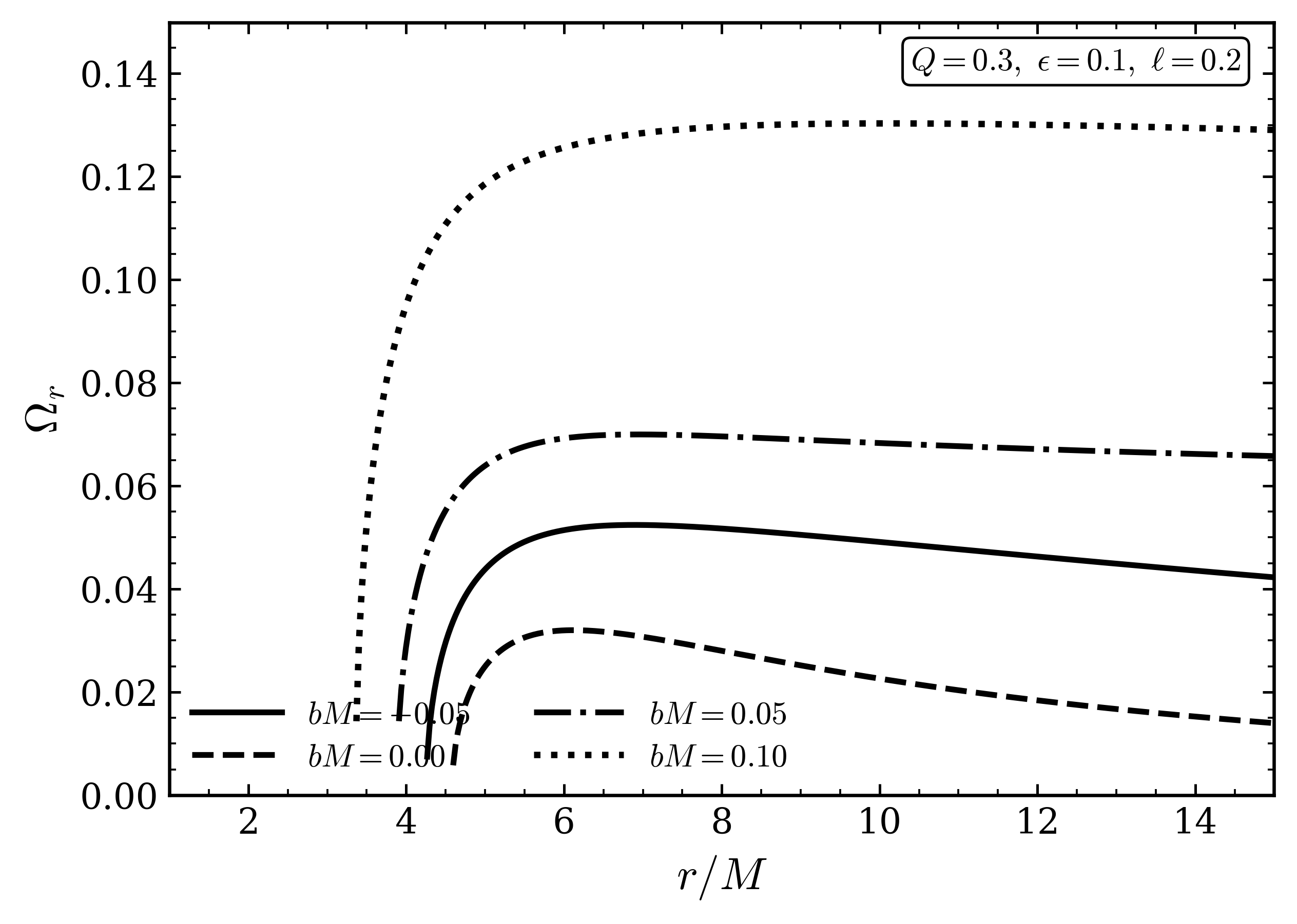}
\caption{Radial epicyclic frequency $\Omega_r$ as a function of radius for
the charged KR black hole. The parameters are fixed as
$M=1$, $Q=0.3M$, $\varepsilon=0.1$, and $\ell=0.2$, while several
values of the magnetic coupling $bM$ are shown. The point where
$\Omega_r$ vanishes marks the onset of marginal radial stability.}
\label{fig:KR_Omega_r}
\end{figure}

Figure~\ref{fig:KR_Omega_r} displays the radial epicyclic frequency, which
governs the oscillation of a particle under small radial displacements from a
circular orbit. As in Fig.~\ref{fig:KR_OmegaK}, the plot label is written
compactly as $\Omega_r$; the physical frequency is
$\Omega_r^{\rm phys}=\sqrt{1-\ell}\,\Omega_r$. Unlike $\Omega_K$, the
radial frequency is not simply a measure of orbital motion; it is a direct
indicator of dynamical stability. When $\Omega_r^2>0$, a radially perturbed
particle executes stable harmonic oscillations around the reference orbit. As
the orbit approaches the strong-gravity region, $\Omega_r$ decreases and
eventually vanishes. This condition is equivalent to the marginal-stability
condition $d^2\mathcal{R}/dr^2=0$ used earlier. Therefore,
$\Omega_r=0$ identifies the ISCO, while the region with imaginary
$\Omega_r$ corresponds to radially unstable circular motion.

The shape of the curves in Fig.~\ref{fig:KR_Omega_r} also shows how the
magnetic coupling modifies the radial stability of the orbit. Since the
magnetic terms in Eq.~\eqref{Omega_r_KR} contain both a linear contribution in
$b\Omega_\phi$ and a quadratic contribution in $b^2\Psi_{\rm KR}'(r)^2$,
the external field affects not only the orbital frequency itself but also the
response of the system to radial perturbations. In physical terms, the
magnetic field changes how efficiently centrifugal support can balance the
combined gravitational and electrostatic attraction. Consequently, the radius
at which $\Omega_r$ goes to zero shifts with $bM$, indicating that the
external magnetic field can move the boundary between stable and unstable
charged-particle orbits.

The vertical epicyclic frequency takes the compact form
\begin{equation}
    \Omega_\theta^2
    =
    \Omega_\phi^2
    +
    \frac{2b\Omega_\phi\Psi_{\rm KR}(r)}
    {r^2u^t}.
    \label{Omega_theta_KR}
\end{equation}
In the absence of magnetic interaction, $b=0$, spherical symmetry gives
\begin{equation}
    \Omega_\theta=\Omega_\phi .
\end{equation}

Since the KR metric is not asymptotically normalized in the usual
Minkowskian form, the physical frequencies are defined with respect to
\[
    t_{\rm phys}=\frac{t}{\sqrt{1-\ell}} .
\]
Thus,
\begin{equation}
    \Omega_i^{\rm phys}
    =
    \sqrt{1-\ell}\,\Omega_i,
    \qquad
    i=(\phi,r,\theta).
    \label{Omega_phys_KR}
\end{equation}
The corresponding frequencies in hertz are
\begin{equation}
    \nu_i
    =
    \frac{1}{2\pi}
    \frac{c^3}{G M_{\rm BH}}
    \sqrt{1-\ell}\,
    \bar{\Omega}_i,
    \qquad
    i=(\phi,r,\theta),
    \label{nu_Hz_KR}
\end{equation}
where $\bar{\Omega}_i$ denotes the dimensionless frequency obtained in
geometrized units.

For the relativistic precession model, the observed QPO frequencies are
identified with the physically normalized frequencies,
\begin{equation}
    \nu_{\rm U}=\nu_\phi,
    \qquad
    \nu_{\rm L}=\nu_\phi-\nu_r .
    \label{RP_KR}
\end{equation}
Equivalently, in angular-frequency notation,
\[
    \Omega_{\rm U}^{\rm phys}=\Omega_K^{\rm phys},
    \qquad
    \Omega_{\rm L}^{\rm phys}
    =
    \Omega_K^{\rm phys}-\Omega_r^{\rm phys}.
\]
These quantities are then compared with the observed upper and lower QPO
frequencies.

The above analysis shows that the KR parameter, the electric charge, and the
external magnetic field all enter the QPO problem through well-defined and
physically distinct mechanisms. The KR deformation modifies the background
geometry and the asymptotic normalization of time, the electric field changes
the force balance through the Coulomb interaction, and the source-free
magnetic field affects both the orbital frequency and the radial stability via
$\Psi_{\rm KR}(r)$, $\Psi'_{\rm KR}(r)$, and
$\Psi''_{\rm KR}(r)$. Therefore, even before fitting any observational data,
the fundamental frequencies already encode a nontrivial fingerprint of the KR
spacetime and its electromagnetic environment. This makes the pair
$(\Omega_K,\Omega_r)$ particularly useful for QPO studies, since after the
common physical-time rescaling these two frequencies directly determine the
relativistic-precession observables $\nu_{\rm U}$ and $\nu_{\rm L}$.

\subsection{Markov chain Monte Carlo analysis for quasi-periodic oscillations}
\label{subsec:mcmc_qpo}

We constrain the charged KR black-hole model using the observed
twin-peak QPO frequencies of GRO~J1655--40, XTE~J1550--564, and M82~X-1.
Within the relativistic precession model, the theoretical upper and lower
frequencies are identified as
\begin{equation}
    \nu_{\rm U}^{\rm th}=\nu_{\phi},
    \qquad
    \nu_{\rm L}^{\rm th}=\nu_{\phi}-\nu_{r},
    \label{eq:mcmc_frequency_identification}
\end{equation}
where $\nu_{\phi}$ and $\nu_r$ are calculated from
Eqs.~\eqref{nu_Hz_KR} and \eqref{RP_KR} at the QPO emission radius $r$.

The observational frequencies used in the likelihood are summarized in
Table~\ref{tab:qpo_observations}. The numerical values correspond to the
QPO pairs adopted in the present analysis, while the cited observational
studies report the corresponding twin-peak detections
\cite{Remillard2002HFQPO,Pasham2014M82}.

\begin{table}[t]
\caption{\label{tab:qpo_observations}
Observed upper and lower QPO frequencies used in the MCMC analysis.
All frequencies are given in hertz.}
\begin{ruledtabular}
\begin{tabular}{lccc}
Source
& $\nu_{\rm U}^{\rm obs}$
& $\nu_{\rm L}^{\rm obs}$
& Reference
\\
\hline
GRO~J1655--40
& $451\pm5$
& $298\pm4$
& \cite{Remillard2002HFQPO}
\\
XTE~J1550--564
& $276\pm3$
& $184\pm5$
& \cite{Remillard2002HFQPO}
\\
M82~X-1
& $5.07\pm0.06$
& $3.32\pm0.06$
& \cite{Pasham2014M82}
\end{tabular}
\end{ruledtabular}
\end{table}

For each source, the observational data vector is written as
\begin{equation}
    \mathcal{D}
    =
    \left\{
        \nu_{\rm U}^{\rm obs},
        \nu_{\rm L}^{\rm obs}
    \right\},
    \qquad
    \boldsymbol{\sigma}
    =
    \left\{
        \sigma_{\rm U},
        \sigma_{\rm L}
    \right\}.
    \label{eq:mcmc_data_vector}
\end{equation}

To keep the notation identical to that used in the corner plots, the
sampled parameter vector is defined as
\begin{equation}
    \boldsymbol{\theta}
    =
    \left(
        \frac{M}{M_{\odot}},
        \frac{Q}{M},
        \ell,
        \epsilon,
        \beta,
        \frac{r}{M}
    \right),
    \label{eq:mcmc_parameter_vector}
\end{equation}
where $M$ is the black-hole mass, $Q/M$ is the dimensionless black-hole
charge, $\ell$ is the Lorentz-violating parameter,
$\epsilon=q/m$ is the particle specific charge,
$\beta=bM=qBM/(2m)$ is the dimensionless magnetic coupling, and $r/M$
is the dimensionless QPO emission radius.

The posterior probability density is given by Bayes' theorem,
\begin{equation}
    \mathcal{P}
    \left(
        \boldsymbol{\theta}
        \mid
        \mathcal{D}
    \right)
    =
    \frac{
        \mathcal{L}
        \left(
            \mathcal{D}
            \mid
            \boldsymbol{\theta}
        \right)
        \pi
        \left(
            \boldsymbol{\theta}
        \right)
    }{
        \mathcal{Z}
        \left(
            \mathcal{D}
        \right)
    },
    \label{eq:mcmc_posterior}
\end{equation}
where $\mathcal{L}$ is the likelihood, $\pi$ is the prior, and
$\mathcal{Z}$ is the Bayesian evidence.

Assuming independent Gaussian errors for the two observed frequencies,
the log-likelihood is
\begin{align}
    \ln \mathcal{L}
    =-\frac{1}{2}
    \Bigg[
    &\left(
        \frac{
            \nu_{\rm U}^{\rm th}
            -
            \nu_{\rm U}^{\rm obs}
        }{
            \sigma_{\rm U}
        }
    \right)^2
    \nonumber\\
    &+
    \left(
        \frac{
            \nu_{\rm L}^{\rm th}
            -
            \nu_{\rm L}^{\rm obs}
        }{
            \sigma_{\rm L}
        }
    \right)^2
    \Bigg].
    \label{eq:mcmc_log_likelihood}
\end{align}

The priors are restricted to the physically allowed region. In
particular, the sampled points must satisfy
\begin{equation}
    \ell<1,
    \qquad
    \left|
        \frac{Q}{M}
    \right|
    <
    (1-\ell)^{3/2},
    \label{eq:mcmc_horizon_constraint}
\end{equation}
together with the conditions that the orbit lies outside the event
horizon, remains timelike, and has a real radial epicyclic frequency.
For Gaussian priors, the logarithm of the prior probability is written as
\begin{equation}
    \ln\pi
    =
    -\frac{1}{2}
    \sum_{i}
    \left(
        \frac{
            \theta_i-\mu_i
        }{
            s_i
        }
    \right)^2,
    \label{eq:mcmc_gaussian_prior}
\end{equation}
inside the imposed parameter intervals, while $\ln\pi=-\infty$ outside
the allowed region.

The posterior distributions were sampled with the affine-invariant
ensemble sampler implemented in the \texttt{emcee} package
\cite{ForemanMackey2013}. For each source, we used 24 walkers. For
GRO~J1655--40 and M82~X-1, the chains were evolved for $10^{4}$ steps,
and the first $2\times10^{3}$ steps were discarded as burn-in. For
XTE~J1550--564, the chains were evolved for $10^{5}$ steps, and the
first $2\times10^{4}$ steps were discarded as burn-in. The remaining
chains were thinned by retaining every fifth sample for GRO~J1655--40
and M82~X-1, and every 25th sample for XTE~J1550--564. The thinning
was used only to reduce the storage size of the posterior samples and
does not affect the sampling procedure itself.

For each parameter, the reported central value is the median of the
marginalized posterior, while the lower and upper uncertainties are
obtained from the 16th and 84th percentiles:
\begin{equation}
    \theta_i
    =
    {\theta_{i,50}}^{+\left(\theta_{i,84}-\theta_{i,50}\right)}
    _{-\left(\theta_{i,50}-\theta_{i,16}\right)} .
    \label{eq:mcmc_percentiles}
\end{equation}

\begin{table*}[t]
\centering
\caption{Gaussian priors adopted in the MCMC analysis. The priors are constructed from a preliminary physically allowed parameter-space scan. Each entry is given in the form $\mu\pm\sigma$.}
\label{tab:mcmc_priors}
\scriptsize
\setlength{\tabcolsep}{4pt}
\begin{tabular}{lcccccc}
\hline
Source & $M/M_\odot$ & $Q/M$ & $\ell$ & $\epsilon$ & $\beta$ & $r/M$ \\
\hline
GRO J1655--40 
& $5.4227 \pm 0.3765$ 
& $0.1854 \pm 0.0839$ 
& $0.0693 \pm 0.0519$ 
& $0.0362 \pm 0.0690$ 
& $-0.0155 \pm 0.0074$ 
& $6.0083 \pm 0.2544$ \\

XTE J1550--564 
& $8.3034 \pm 0.3954$ 
& $0.2330 \pm 0.0807$ 
& $0.1302 \pm 0.0324$ 
& $0.1676 \pm 0.0478$ 
& $-0.0051 \pm 0.0065$ 
& $5.6544 \pm 0.1969$ \\

M82 X-1 
& $519.6918 \pm 36.9433$ 
& $0.2226 \pm 0.0819$ 
& $0.0927 \pm 0.0478$ 
& $-0.1569 \pm 0.0535$ 
& $-0.0199 \pm 0.0072$ 
& $5.8073 \pm 0.2363$ \\
\hline
\end{tabular}
\end{table*}

The MCMC analysis was performed using Gaussian priors for the six model
parameters. The mean values and standard deviations of these priors are
listed in Table~\ref{tab:mcmc_priors}. The priors were obtained from a
preliminary scan of the physically admissible parameter space, where the
horizon condition, timelike circular motion, and real epicyclic
frequencies were imposed. Since the model contains six free parameters,
whereas each source provides two observed QPO frequencies, the adopted
priors help restrict the sampling to physically meaningful regions of
the parameter space. Consequently, the posterior constraints should be
interpreted within the adopted model and prior assumptions.

Because the posterior constraints can be sensitive to the adopted priors, especially in a multi-parameter model constrained by two QPO frequencies, the prior choices are explicitly reported for transparency.

\begin{figure}[t]
\centering
\includegraphics[width=0.98\linewidth]{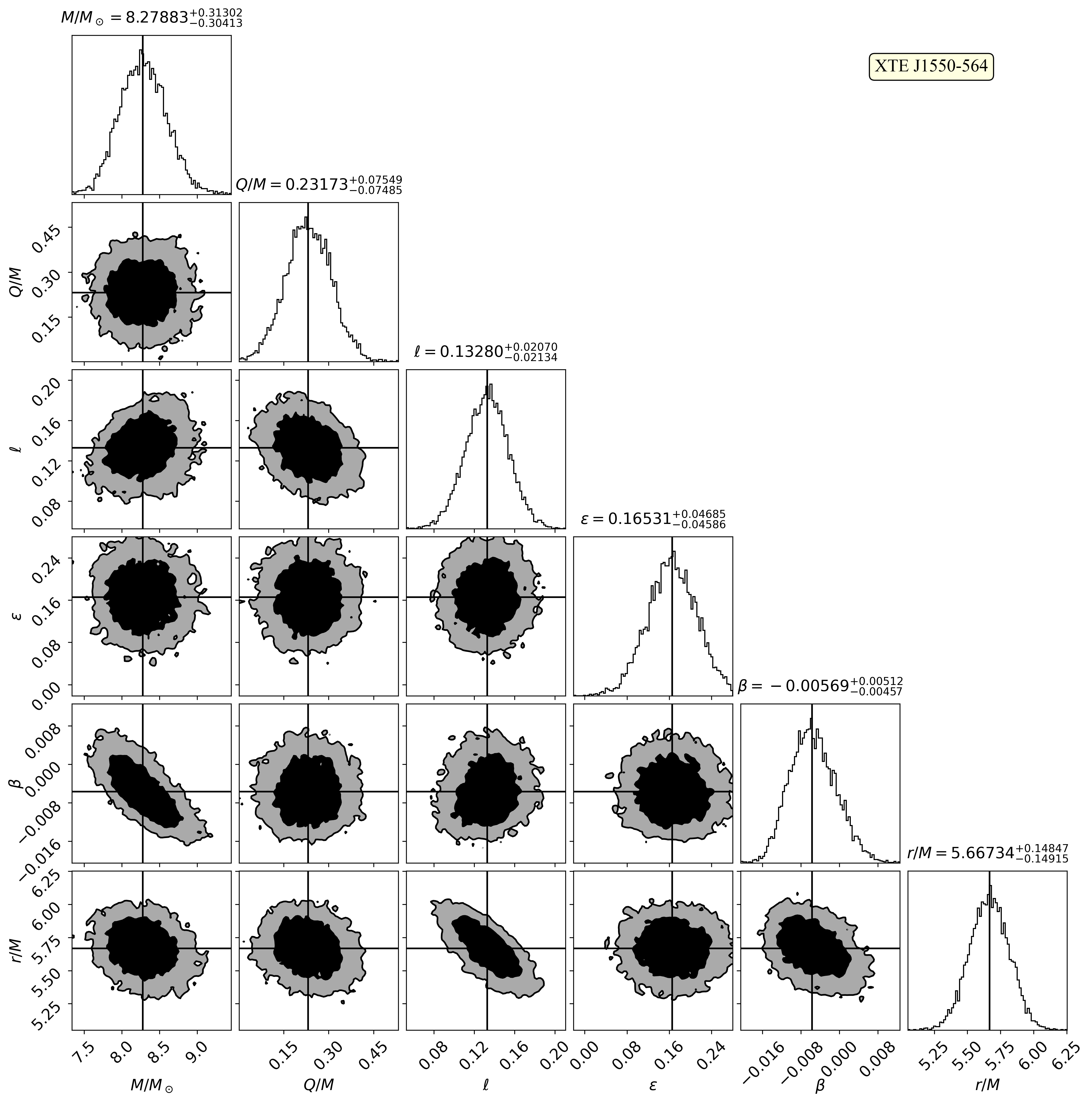}
\caption{\label{fig:xte_corner}
Marginalized posterior distributions of
$M/M_{\odot}$, $Q/M$, $\ell$, $\epsilon$, $\beta$, and $r/M$
for XTE~J1550--564.}
\end{figure}

\begin{figure}[t]
\centering
\includegraphics[width=0.98\linewidth]{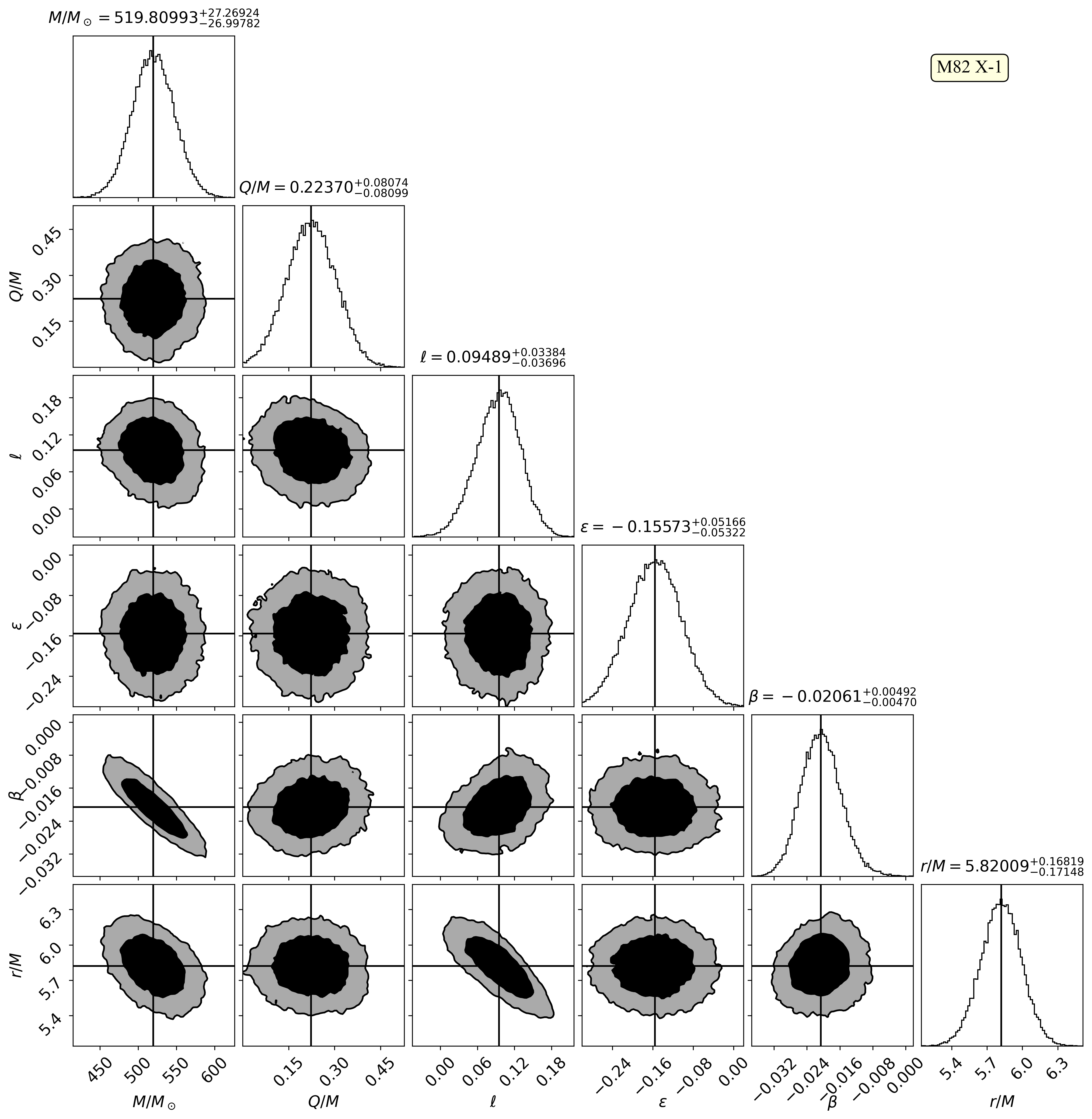}
\caption{\label{fig:m82_corner}
Marginalized posterior distributions of
$M/M_{\odot}$, $Q/M$, $\ell$, $\epsilon$, $\beta$, and $r/M$
for M82~X-1.}
\end{figure}

\begin{figure}[t]
\centering
\includegraphics[width=0.98\linewidth]{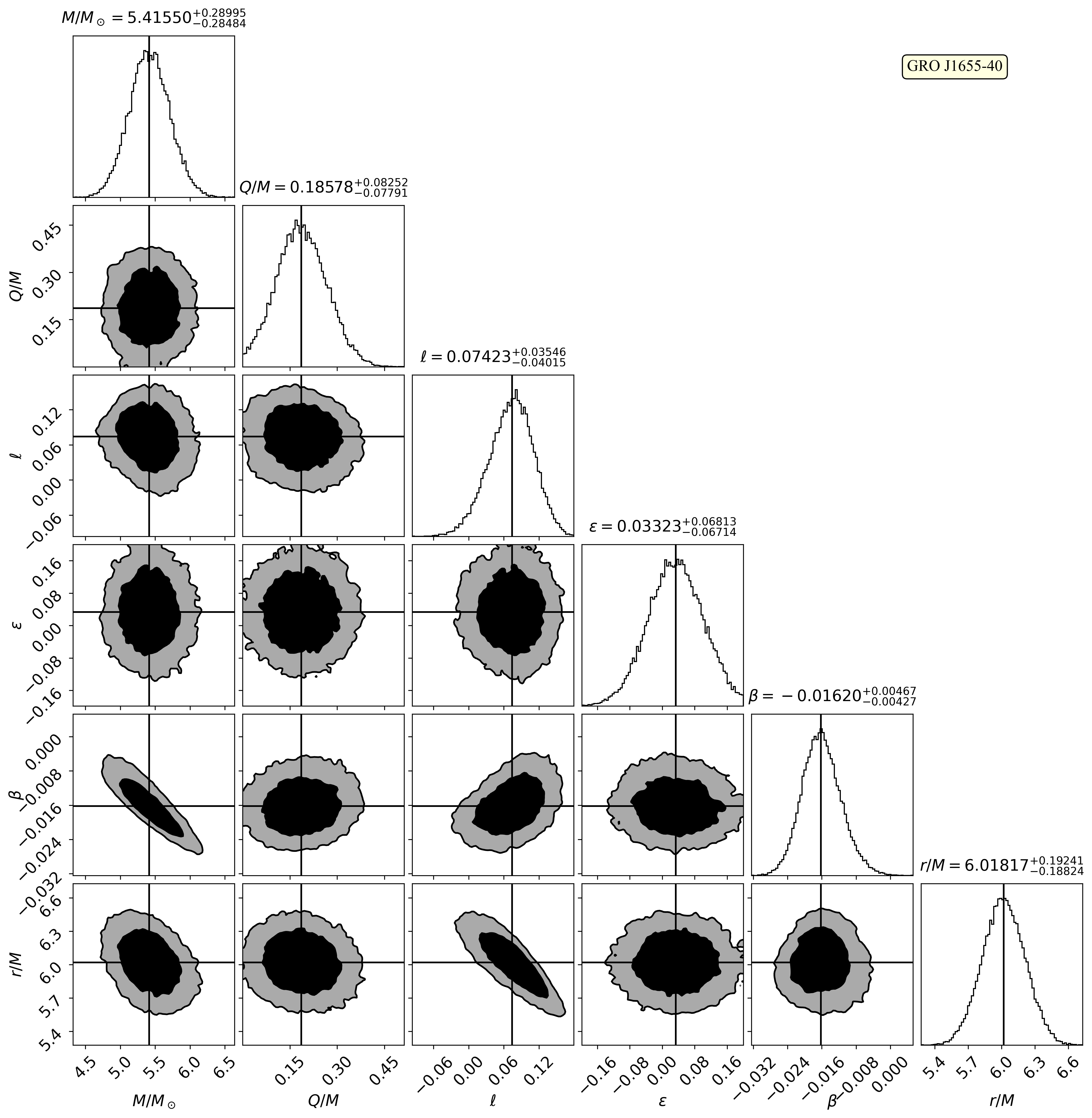}
\caption{\label{fig:gro_corner}
Marginalized posterior distributions of
$M/M_{\odot}$, $Q/M$, $\ell$, $\epsilon$, $\beta$, and $r/M$
for GRO~J1655--40.}
\end{figure}

The marginalized posterior constraints obtained from the MCMC samples
are summarized in Table~\ref{tab:mcmc_posteriors}, while the
corresponding one- and two-dimensional posterior distributions are
shown in Figs.~\ref{fig:xte_corner}, \ref{fig:m82_corner}, and
\ref{fig:gro_corner}.

\begin{table}[t]
\caption{\label{tab:mcmc_posteriors}
Marginalized posterior constraints. The uncertainties correspond to the 16th and 84th percentiles of the marginalized posterior samples.}
\begin{ruledtabular}
\begin{tabular}{lccc}
Parameter
& GRO~J1655--40
& XTE~J1550--564
& M82~X-1
\\
\hline
$M/M_{\odot}$
& $5.41550^{+0.28995}_{-0.28484}$
& $8.27883^{+0.31302}_{-0.30413}$
& $519.80993^{+27.26924}_{-26.99782}$
\\
$Q/M$
& $0.18578^{+0.08252}_{-0.07791}$
& $0.23173^{+0.07549}_{-0.07485}$
& $0.22370^{+0.08074}_{-0.08099}$
\\
$\ell$
& $0.07423^{+0.03546}_{-0.04015}$
& $0.13280^{+0.02070}_{-0.02134}$
& $0.09489^{+0.03384}_{-0.03696}$
\\
$\epsilon$
& $0.03323^{+0.06813}_{-0.06714}$
& $0.16531^{+0.04685}_{-0.04586}$
& $-0.15573^{+0.05166}_{-0.05322}$
\\
$\beta$
& $-0.01620^{+0.00467}_{-0.00427}$
& $-0.00569^{+0.00512}_{-0.00457}$
& $-0.02061^{+0.00492}_{-0.00470}$
\\
$r/M$
& $6.01817^{+0.19241}_{-0.18824}$
& $5.66734^{+0.14847}_{-0.14915}$
& $5.82009^{+0.16819}_{-0.17148}$
\end{tabular}
\end{ruledtabular}
\end{table}

Figures~\ref{fig:xte_corner}--\ref{fig:gro_corner} display the
one-dimensional marginalized distributions along the diagonal and the
pairwise joint posterior distributions in the off-diagonal panels. The
vertical and horizontal lines indicate the posterior median values.

\begin{table}[t]
\caption{\label{tab:mcmc_diagnostics}
MCMC convergence diagnostics for the three QPO sources. The quantity
$\tau_{\rm int}$ denotes the range of integrated autocorrelation times
over the six model parameters.}
\begin{ruledtabular}
\begin{tabular}{lcccc}
Source & Acceptance fraction & $\tau_{\rm int}$ range & $N_{\rm step}/\tau_{\rm max}$ & Final samples \\
\hline
GRO~J1655--40 
& 0.4807 
& 73.87--110.51 
& 90.49 
& 38400 \\
XTE~J1550--564 
& 0.5043 
& 65.11--84.00 
& 1190.48 
& 76800 \\
M82~X-1 
& 0.5025 
& 68.77--82.70 
& 120.92 
& 38400 \\
\end{tabular}
\end{ruledtabular}
\end{table}

The convergence of the chains was checked using the mean acceptance
fraction and the integrated autocorrelation time. As shown in
Table~\ref{tab:mcmc_diagnostics}, the mean acceptance fractions lie in
the range $0.48$--$0.50$, indicating efficient sampling. The ratio
$N_{\rm step}/\tau_{\rm max}$ is larger than $90$ for all sources,
showing that the chains are sufficiently longer than the corresponding
autocorrelation times. Therefore, the marginalized posterior
constraints reported in Table~\ref{tab:mcmc_posteriors} can be
estimated reliably within the adopted model and prior ranges.

The corner plots reveal several parameter correlations. In particular,
the magnetic coupling $\beta$ is correlated with the mass and the
orbital radius, while $\ell$ and $r/M$ exhibit an appreciable
anti-correlation for XTE~J1550--564 and GRO~J1655--40. These
correlations reflect the fact that the observed QPO pair constrains
combinations of the model parameters more strongly than each parameter
separately. Consequently, the quoted intervals should be interpreted
within the adopted six-parameter model and prior ranges.

\begin{table}[t]
\caption{\label{tab:qpo_bestfit}
Comparison between the observed and best-fit theoretical QPO frequencies.
All frequencies are given in Hz.}
\begin{ruledtabular}
\begin{tabular}{lccccc}
Source & $\nu_U^{\rm obs}$ & $\nu_U^{\rm th}$ & $\nu_L^{\rm obs}$ & $\nu_L^{\rm th}$ & $\chi^2_{\rm min}$ \\
\hline
GRO~J1655--40 
& 451.00 
& 450.9703 
& 298.00 
& 298.0015 
& $3.55\times10^{-5}$ \\

XTE~J1550--564 
& 276.00 
& 276.0261 
& 184.00 
& 183.8560 
& $9.05\times10^{-4}$ \\

M82~X-1 
& 5.07 
& 5.0689 
& 3.32 
& 3.3178 
& $1.64\times10^{-3}$ \\
\end{tabular}
\end{ruledtabular}
\end{table}

Table~\ref{tab:qpo_bestfit} compares the observed QPO frequencies with
the best-fit theoretical values obtained in the present model. The very
small residuals in Table~\ref{tab:qpo_bestfit} show that the model can
reproduce the adopted QPO pairs at the best-fit points. Since the model
contains several free parameters, these small values should be viewed as
a consistency check of the frequency reconstruction rather than as an
independent model-selection criterion.

\clearpage

\section{Conclusion}\label{sec:conclusion}

In this work, we investigated charged-particle dynamics and quasi-periodic oscillations around a Reissner--Nordstr\"om-like black hole in KR gravity in the presence of an external magnetic test field. The KR background introduces a Lorentz-violating parameter, which modifies the metric function, the horizon structure, the circular-orbit properties, and the characteristic frequencies of particle motion. A central point of the present analysis is that the magnetic field was not imposed by using the standard Wald-type prescription. Instead, it was obtained from the source-free Maxwell equation on the charged KR background. This makes the magnetic configuration consistent with the modified geometry and its asymptotic structure.

We derived the equations of motion for charged test particles, constructed the effective potential, and analyzed the conditions for circular orbits and their stability. The results show that the black-hole charge, the KR parameter, the specific charge of the particle, and the magnetic coupling jointly determine the behavior of charged-particle motion. The magnetic field modifies the mechanical angular momentum of the particle, while the KR parameter changes the gravitational background itself. Therefore, the ISCO and the orbital frequency are affected by the combined influence of the geometrical deformation and the electromagnetic interaction. The ISCO results summarized in Table~\ref{tab:kr_isco_quantities} show that the stable-orbit region is sensitive to both the KR deformation and the magnetic coupling, indicating that electromagnetic effects may remain important even in the test-field approximation.

We also derived the orbital and radial epicyclic frequencies and applied them to high-frequency QPOs within the relativistic precession model. In this framework, the upper QPO frequency is associated with the orbital frequency, while the lower QPO frequency is related to the periastron-precession frequency. Since these frequencies are generated in the strong-field region close to the compact object, they provide a useful phenomenological probe of the spacetime geometry and the surrounding electromagnetic environment.

The observed QPO data used in the statistical analysis are listed in Table~\ref{tab:qpo_observations}. These data were used for the three sources GRO~J1655--40, XTE~J1550--564, and M82~X-1. The model parameters were constrained by means of a Markov chain Monte Carlo analysis. The adopted Gaussian priors are presented in Table~\ref{tab:mcmc_priors}, while the corresponding posterior constraints are summarized in Table~\ref{tab:mcmc_posteriors}. This structure makes the statistical analysis transparent: the observational input is given separately, the prior assumptions are explicitly stated, and the final parameter constraints are reported independently in the posterior table.

According to the posterior constraints in Table~\ref{tab:mcmc_posteriors}, the charged KR black-hole model with an external magnetic test field can reproduce the observed QPO pairs of all three sources within the adopted model and prior ranges. The inferred masses remain consistent with the expected astrophysical character of the corresponding compact objects. The posterior distributions also indicate that the QPO data are compatible with nonzero values of the charge and KR parameters in the considered framework. In addition, the preferred emission radii reported in Table~\ref{tab:mcmc_posteriors} lie in the strong-field region, where the orbital and radial epicyclic frequencies are particularly sensitive to changes in the background geometry.

The convergence diagnostics reported in Table~\ref{tab:mcmc_diagnostics} support the reliability of the sampling procedure. The acceptance fractions and integrated autocorrelation times show that the chains sample the posterior distributions efficiently for the three sources. The corresponding corner plots illustrate the marginalized posterior distributions and the correlations among the model parameters. These correlations are physically expected, because the observed QPO frequencies depend simultaneously on the mass, charge, KR parameter, magnetic coupling, specific particle charge, and emission radius.

Overall, the present results suggest that QPO observations can provide meaningful phenomenological constraints on Lorentz-violating black-hole spacetimes when the effects of charge and magnetic fields are included. The analysis also shows that the KR deformation and the magnetic interaction should not be treated as completely independent observational effects, since both can influence the same strong-field frequencies. Thus, the combined study of particle dynamics, ISCO behavior, and QPO frequencies offers a useful way to test modified black-hole geometries in astrophysical environments.

At the same time, the obtained constraints should be interpreted with appropriate care. The model contains several free parameters, whereas each source provides only a limited number of observed QPO frequencies. Therefore, the posterior constraints depend on the adopted priors and on the assumptions of the relativistic precession model. Moreover, the magnetic field was treated in the test-field approximation, and the effects of plasma dynamics, detailed accretion-disk structure, radiation transport, magnetic-field backreaction, and black-hole spin were not included.

Future work may extend the present analysis to rotating KR black holes and to more realistic disk and plasma models. It would also be useful to combine QPO constraints with other observables, such as black-hole shadows, continuum spectra, and iron-line profiles. Such a combined approach may help reduce parameter degeneracies and provide stronger observational tests of Lorentz-violating gravity in the strong-field regime.

\section*{Acknowledgments}
S.M. gratefully acknowledges support from Grant FZ-20200929385 of the Agency of Innovative Developments of the Republic of Uzbekistan.

\bibliographystyle{apsrev4-2}
\bibliography{references}

\end{document}